\documentclass[twocolumn,aps,showpacs,prb,tightenlines,amsmath,amssymb]{revtex4}
\usepackage{graphicx}
\usepackage{amssymb}
\usepackage{slashed}
\usepackage{dcolumn}
\usepackage{amsmath}
\usepackage{bm}
\usepackage{colordvi}
\usepackage{mathrsfs}
\makeatletter

\newcommand{\Rmnum}[1]{\expandafter\@slowromancap\romannumeral #1@}
\makeatother

\begin{document}

\title{Gauge-invariant microscopic kinetic theory of superconductivity in
  response to electromagnetic fields}   

\author{F. Yang}
\affiliation{Hefei National Laboratory for Physical Sciences at
Microscale, Department of Physics, and CAS Key Laboratory of Strongly-Coupled
Quantum Matter Physics, University of Science and Technology of China, Hefei,
Anhui, 230026, China}

\author{M. W. Wu}
\thanks{Author to whom correspondence should be addressed}
\email{mwwu@ustc.edu.cn.}

\affiliation{Hefei National Laboratory for Physical Sciences at
Microscale, Department of Physics, and CAS Key Laboratory of Strongly-Coupled
Quantum Matter Physics, University of Science and Technology of China, Hefei,
Anhui, 230026, China}

\date{\today}

\begin{abstract} 

Within a gauge-invariant microscopic kinetic theory, we study the
electromagnetic response in the superconducting states. Both superfluid and 
normal-fluid dynamics are involved. We predict that the normal fluid is present
only when the excited superconducting velocity $v_s$ is larger than a threshold
$v_L=|\Delta|/k_F$. Interestingly, with the normal
fluid, we find that there exists friction between the normal-fluid and superfluid
currents. Due to this friction, part of the superfluid becomes
viscous. Therefore, a three-fluid model: normal fluid, non-viscous and
viscous superfluids, is proposed. For the stationary magnetic response, at
$v_s<v_L$ with only the non-viscous superfluid, the Meissner 
supercurrent is excited and the gap equation can reduce to Ginzburg-Landau equation.
At $v_s{\ge}v_L$, with the normal fluid, non-viscous and viscous
superfluids, in addition to the directly excited Meissner supercurrent in the
superfluid, normal-fluid current is also induced through the friction 
drag with the viscous superfluid current. Due to the normal-fluid and
viscous superfluid currents, the penetration depth is influenced by the
scattering effect. In addition, a modified Ginzburg-Landau
equation is proposed. We predict an exotic 
phase in which both the resistivity and superconducting gap are {\em
  finite}. As for the optical response, the excited ${v_s}$ 
oscillates with time. When $v_s<v_L$, only the non-viscous superfluid is present
whereas at 
$v_s{\ge}v_L$, normal fluid, non-viscous and viscous superfluids are present. 
We show that the excited normal-fluid current exhibits the Drude-model
behavior while the superfluid current consists of the Meissner supercurrent 
and Bogoliubov quasiparticle current. Due to the friction between the
superfluid and normal-fluid currents, the optical conductivity is captured by the
three-fluid model. Finally, we also study the optical excitation of the Higgs
mode. By comparing the contributions from the drive and Anderson-pseudospin pump
effects, we find that the drive effect is dominant at finite temperature whereas
at zero temperature, both effects contribute. 

\end{abstract}


\maketitle 

\section{Introduction}

In the field of superconductivity, electromagnetic responses have been
attracting intensive attention in the past few decades for revealing
the physics of superconductivity and exploring the novel
properties.\cite{Ba0,Ba1,Ba2,Ba3,Ba4,Ba5,Ba6,Ba7,Ba8}  
For the stationary magnetic response, the induced diamagnetic
supercurrent and the resulting magnetic-flux expulsion are known to be one of
the fundamental phenomena in superconductors, referred to as Meissner
effect.\cite{Meissner,London} Analysis of the magnetic response
in the early-stage works are based on the well-known
Ginzburg-Landau phenomenological theory for pure superconductors.\cite{Ginzburg}    
As for the optical studies in superconductors, efforts are focused on the
microwave and terahertz (THz) absorptions in both
linear\cite{L1,L2,L3,L4,L5,L6,L7,L8,L9} and 
nonlinear\cite{NL1,NL2,NL3,NL4,NL5,NL6,NL7,NL8,NL9,NL10} regimes. Particularly, a
phenomenological picture based on the two fluid model, 
which was first proposed by Tisza and London\cite{Tisza} and then developed by 
Landau\cite{Landau} in bosonic liquid helium II, is widely used
to capture the physics of the optical response in 
superconductors.\cite{Ba0,Ba1,Ba4,L6,L7,L9,NL1,NL2,two} It is postulated
that both the normal fluid and superfluid are present as separate fluids, each with
its own density and velocity in the superconducting
state. The normal fluid in the optical response exhibits the Drude-model
behaviors.\cite{Ba0,Ba1,Ba4,L6,L7,L9,NL1,NL2} Superfluid on the other hand has 
no resistivity.\cite{Ba0,Ba1,Ba4,L6,L7,L9,NL1,NL2} Recently, it was
experimentally realized that through the intense THz field,  
one can excite the fluctuation of the superfluid density with the oscillation
frequency at twice optical frequency.\cite{NL5,NL7,NL8,NL9,NL10} This
oscillation so far is attributed to the excited Higgs mode, i.e.,   
fluctuation of the magnitude of the superconducting order
parameter.\cite{Higgs1,Higgs2,Higgs20,Higgs21,Higgs3,Higgs4,Higgs5,Higgs50,Higgs6,Higgs7,Higgs8,Higgs9,Higgs10}   
In most situations, a plateau of the superconducting order parameter is
discovered after the THz pulse.\cite{NL7,NL8} 

Within the framework of superconductivity theory established by Bardeen, Cooper,
and Schrieffer (BCS),\cite{BCS} microscopic theories of the above
electromagnetic properties of superconductors have been developed for more than five
decades.\cite{Higgs20,Higgs21,Higgs4,Higgs5,Higgs50,Higgs6,Higgs7,Higgs8,Higgs10,MB,MBm,Gor-G,Gor-A,Gor-o,G1,Gor-oMB,G2,MBo,Eilen,Usadel,GOBE1,GOBE2,GOBE3}   
In principle, a complete theory to calculate the electromagnetic 
properties must satisfy certain conditions. First, it should be capable of
calculating both magnetic and optical responses in linear and
nonlinear regimes. Second, it should include the scattering effect, which is
inevitable in dirty superconducting metals.\cite{Higgs4,Higgs5} Finally, it should
satisfy the gauge invariance in superconductors,\cite{gi0,gi1,gi2} first revealed
by Nambu\cite{gi0,gi2} based on a gauge structure of vector potential ${\bf A}$, scalar
potential $\phi$ and superconducting phase $\psi$. However, to the best of our
knowledge, a microscopic theoretical description which satisfies all three conditions above,
is still absent in the literature.  

Specifically, the electromagnetic properties of  
conventional superconductors was first discussed by Mattis and Bardeen (MB)
within the BCS theory in the
linear regime and dirty limit.\cite{MB}  Based on the MB theory, Miller gave a
dependence of the penetration depth $\delta$ on mean free path $l$ in the case of a
stationary magnetic response.\cite{MBm} This dependence was extended by Tinkham to 
the regime between clean and dirty limits later as
$\delta=\delta_{c}\sqrt{1+{\xi_0}/{l}}$ at low temperature\cite{Ba1} ($\xi_0$
and $\delta_{c}$ denote the coherence length and clean-limit penetration depth,
respectively), in good agreement with the 
experiments.\cite{E1,E2,E3,E4,E5} This directly indicates that the Meissner
supercurrent experiences a friction resistance from
scattering. Nevertheless, a supercurrent should be non-viscous. 
The physical origin of the friction 
resistance on a supercurrent is still unclear in the literature, since the scattering effect in
the early-stage works\cite{MB,Ba1} is included through a hand-waving discussion
and hence the microscopic scattering process is absent. 
As for the optical response, MB theory reveals that the optical absorption is
realized by breaking the Cooper pairs into the quasielectrons and quasiholes when 
the optical frequency is larger than twice the superconducting-gap
magnitude.\cite{MB,Ba0,MBo} In this regime, the MB theory successfully describes 
the experimentally observed complex conductivity.\cite{L3,L4,L5,L7,NL7}
However, at low frequency, it deviates from the experimental
observation.\cite{L3,L5,NL7} In addition to this deficiency, it is hard to 
extend the MB theory into the nonlinear regime, and hence, the 
excitation of Higgs mode is absent in this description. Most importantly,
as an early-stage work, the MB theory,\cite{MB} established in a
specific gauge with finite vector potential alone, is not gauge invariant.      

Theories for the excitation of the Higgs mode in superconductors are mostly based 
on the Liouville\cite{Higgs20,Higgs21,Higgs5} or Bloch\cite{Higgs4,Higgs50,Higgs6,Higgs7,Higgs8,Higgs10}
equation derived in the Anderson  
pseudospin representation.\cite{As} In these theories, the
nonlinear term ${\bf 
  A}^2$ is included, which leads to the pump of the quasiparticle correlation
(pump effect) and then contributes to the excitation of the Higgs mode.
However, no drive effect (linear term) is included in this description.
Thus, unphysical conclusions are immediately obtained. On one hand, no optical
current is excited. On other hand, the elastic scattering is ineffective since
the pump effect alone is isotropic in the momentum space.
Consequently, the Liouville\cite{Higgs20,Higgs21,Higgs5} or
Bloch\cite{Higgs4,Higgs50,Higgs6,Higgs7,Higgs8,Higgs10} equation in the literature is insufficient to
elucidate the complete physics. Moreover,
with only finite vector potential,\cite{Higgs20,Higgs21,Higgs4,Higgs5,Higgs50,Higgs6,Higgs7,Higgs8,Higgs10} the
gauge invariance is also unsatisfied.

To data, the most effective method of calculating the electromagnetic properties
in superconductors is provided by Gorkov's equation of Green
function\cite{Gor-G,Gor-A,Gor-o,G1,G2} and its derivatives. Specifically, in the Gorkov's
equation, the gauge invariance is satisfied. For the  
stationary magnetic response, it is demonstrated that the Gorkov's equation can
reduce to the Ginzburg-Landau theory.\cite{Gor-G,G1} 
Moreover, by calculating the scattering self-energy via assuming that the
scattering in superconductors is same as that in normal metals, the disorder
effect on penetration depth is discussed by Abrikosov and
Gorkov,\cite{Gor-A} in consistency with the MB theory.\cite{MBm} 
As for the optical case, it is reported that in appropriate 
limits, the obtained optical conductivity from the Gorkov's equation can reduce to
the MB theory in the dirty 
limit\cite{Gor-oMB} and exhibits the two-fluid-model behavior in the weak
scattering.\cite{Gor-o} However, the Gorkov's equation\cite{G1,G2}
actually is very hard to handle for a kinetic calculation of the temporal evolution or
spatial diffusion in superconducting systems as too many variables are involved.  
The complex calculation also makes it difficult to explore the microscopic
process and physical picture of both the electromagnetic properties and
scattering effect.   

To reduce the number of variables, two kinds of the transformations of Gorkov's
equation into the transport-like equation are developed in the
literature. Specifically, based on Gorkov's equation,
via $\tau_3$-Green function 
[{\small $G(x,x')=-i\tau_3\langle{T}\Psi(x)\Psi^{\dagger}(x')\rangle$} with
{\small $\Psi(x)$} being the Nambu-space field operator,\cite{G1,G2} $x=(t,{\bf
  r})$ denoting the time-space point, $T$ being the chronological ordering\cite{G1} and
$\langle...\rangle$ representing the ensemble average],  
in the quasiclassical approximation\cite{Ba5,Ba6,GQ1} with 
an integration over the energy variable,\cite{GQ1} 
Eilenberger derived a transport-like
equation\cite{Eilen} which can reduce to Ginzburg-Landau equation near the
critical temperature.\cite{Eilen2Ginz} However, the Gauge invariance is lost
during this  
derivation. It is fixed years later\cite{EG1,EG2} by constructing the
gauge-invariant $\tau_3$-Green function via introducing the Wilson
line.\cite{Wilson}  The Eilenberger equation successfully describes the
topics like Josephson effect  
in multilayer junctions,\cite{EJ1,EJ2,EJ3} 
unconventional superconductivity,\cite{EU1,EU2,EU3,EU4} vortex
behaviors\cite{EV1,EV2,EV3,EV4} and disorder influence on 
superconductivity.\cite{ED1,ED2,ED3,ED4} Particularly, for the stationary case
in dirty limit, the Eilenberger equation is further simplified into a
diffusive Usadel equation,\cite{Usadel} 
which is widely used to investigate the superconducting proximity effects in
multilayered structures.\cite{Ba7,Ba8,U1,U2,U3,U4,U5,U6}
However, the specific scattering term in the
Eilenberger equation is very hard to handle due to the relative-time (i.e.,
frequency) variable. Thus, the relaxation-time approximation is usually
taken. Therefore, the microscopic process and physical picture of the scattering
effect are lacking. Moreover, the relative-time variable also
markedly enlarges the difficulty for the temporal evolution. Consequently, it is
hard to apply the Eilenberger equation in the optical study.

Actually, in the optics\cite{GQ2} and spintronics\cite{GQ3} of semiconductors, to
obtain the kinetic equation, a complete nonequilibrium approach with reduced
relative-time variable by taking the equal-time approximation, has been well established.  
Similarly, considering the fact that the superconductivity in
conventional superconductors is characterized by equal-time
pairing,\cite{BCS} Yu and Wu proposed another transformation
of the Gorkov's equation into the transport-like equation in
superconducting states through $\tau_0$-Green function [{\small
  $G(x,x')=-i\langle{T}\Psi(x)\Psi^{\dagger}(x')\rangle$}].\cite{GOBE1}    
Moreover, to retain the gauge invariance, a gauge-invariant
$\tau_0$-Green function\cite{GOBE1} is constructed by introducing the Wilson 
line.\cite{Wilson} Then, a gauge-invariant kinetic equation is proposed.
Thanks to the reduced relative-time variable, this equation is much easier to
handle for the temporal evolution and hence the optical response in
superconductors. Moreover, due to its gauge invariance, both the drive and pump effects 
mentioned above are kept. Particularly, it is revealed that the drive effect
makes a dominant contribution in the Higgs-mode excitation,\cite{GOBE1} in sharp
contrast to the conclusion by Liouville\cite{Higgs20,Higgs21,Higgs5} or Bloch\cite{Higgs4,Higgs50,Higgs6,Higgs7,Higgs8,Higgs10}
equation in which only the pump effect is considered. Most importantly, the
complete microscopic scattering 
process is constructed in this gauge-invariant theory, and the rich
physics of the relaxation mechanism\cite{GOBE1} and transport
phenomena\cite{GOBE3} is revealed. The experimentally observed plateau of the 
superconducting gap after the THz pulse\cite{NL7,NL8} is also
revealed as the consequence of the scattering effect.\cite{GOBE1} 
However, in spite of the success in optical studies, as a
gauge-invariant work for the electromagnetic response, 
this theory fails to apply to the magnetic  
case since it is incapable of giving the Meissner current and reducing to the
Ginzburg-Landau theory.  Therefore, it is natural to 
conclude that this theory only describes the dynamics of
quasiparticles.\cite{GOBE1,GOBE3} Dynamics of
superfluid is not directly involved in this description, but circumvented
through the response of the gap in the Bogoliubov quasiparticle excitation.   

In this work,  we extend the kinetic theory by Yu and
Wu\cite{GOBE1} to include the superfluid, so that both normal-fluid and
superfluid dynamics are involved in the theory. 
As a gauge-invariant theory for the electromagnetic response, our kinetic
equation can be applied to study both the magnetic and optical
cases. We first focus on the weak-scattering case in the 
present work. Rich physics is revealed. Specifically, in the electromagnetic
response, we show that the superconducting 
velocity $v_s$ is always excited. Particularly, 
a threshold $v_L=|\Delta|/k_F$
($\Delta$ and $k_F$ denote the superconducting order parameter and Fermi
momentum, respectively) of superconducting velocity for the emergence of the
normal fluid and hence the scattering is predicted from our theory, i.e., the normal
fluid is excited only when $v_s>v_L$. Actually, similar threshold for the
emergence of the normal fluid and scattering was first proposed by Landau
to interpret the fluid viscosity in bosonic 
liquid helium II at large velocity.\cite{Landau}  Therefore, we refer to this 
threshold as Landau threshold.
Interestingly, we find that there also exists friction between the normal-fluid
and superfluid currents. Due to this friction, part of superfluid
becomes viscous. Therefore, the superfluid consists of the non-viscous
superfluid and viscous one. Consequently, to capture the physics of the
electromagnetic response in superconducting states, a three-fluid model at
$v_s{\ge}v_L$ is proposed from our theory: normal fluid, non-viscous and viscous
superfluids. 

\begin{figure}[htb]
  {\includegraphics[width=8.0cm]{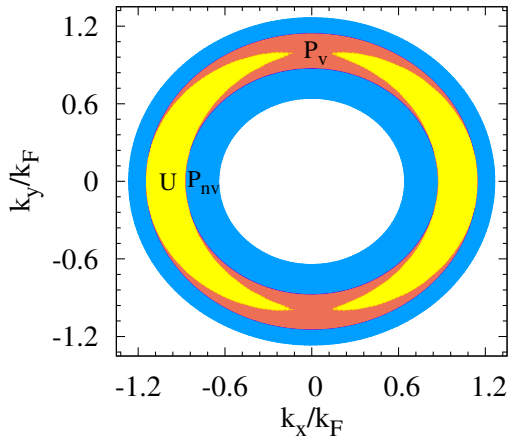}}
\caption{Schematic showing the division in the momentum space when the
  superconducting velocity $v_s$ is larger than 
  the Landau threshold $v_L$. In the figure, the spherical
  shell by the BCS theory is divided into three parts:
  unpairing (U) region characterized by $|{\bf k}\cdot{\bf v}_s|>E_k$, denoted
  by yellow regions; non-viscous pairing (P$_{\rm nv}$) region characterized 
  by $kv_s<E_k$, denoted by purple regions;
  viscous pairing (P$_{\rm v}$) region characterized by $kv_s>E_k$ and $|{\bf
    k}\cdot{{\bf v}_s}|<E_k$, denoted by blue
  regions. }   
\label{figyw1}
\end{figure}

The physics behind these predictions can be understood as follows. It is
established\cite{FF4,FF5,FF6,FF7,FF8,FF9} that with a superconducting
velocity, the quasiparticle energy spectrum is tilted as 
$E^{\pm}_{\bf k}={\bf k}\cdot{\bf v}_s\pm{E_k}$ with $E^{+}_{\bf k}$
($E^{-}_{\bf k}$) standing for the quasielectron (quasihole) energy and $E_k$
being the BCS Bogoliubov quasiparticle energy. At a small superconducting
velocity, the superconducting state behaves like the
BCS state, in which all particles in the spherical shell by the BCS theory
participate in the pairing. Thus, there only exists superfluid.
As for the case with a large superconducting velocity at $v_s{\ge}v_L$, 
in addition to the pairing (P) region with  $|{\bf k}\cdot{\bf v}_s|<E_k$,
there also exists the region with $|{\bf k}\cdot{\bf v}_s|>E_k$, in which 
the quasielectron energy $E^{+}_{\bf k}$ is smaller than zero or the quasihole
energy $E^{-}_{\bf k}$ is larger than zero.  As revealed in the previous
works,\cite{FF7,FF8,FF9,FF1}  the anomalous correlation in this region is
destroyed. Thus, particles in this
regions no longer participate in the pairing and behave like the normal
ones. Following the terminology in the Fulde-Ferrell-Larkin-Ovchinnikov (FFLO)
state,\cite{FF1,FF2} this region is referred to as the unpairing (U) region.
Then, both the normal fluid (from U region) and superfluid (from P
region) are present. Particularly, as shown in Fig.~\ref{figyw1}, there exists a  
special region (P$_{\rm v}$ region characterized by $kv_s>E_k$ and $|{\bf
    k}\cdot{{\bf v}_s}|<E_k$) in the pairing region which shares the 
same momentum magnitude with U region. In conventional
superconducting metals, due to the strong screening, the impurity scattering
behaves as the short-range impurity scattering, which is isotropic in the
momentum space. Therefore, the particles in P$_{\rm v}$ region participate in the
pairing but experience
the scattering with those in U region, leading to the friction between the
superfluid and normal-fluid currents. Consequently, the superfluid in
P$_{\rm v}$ region becomes viscous. Whereas the superfluid in the remaining
pairing region (P$_{\rm nv}$ region characterized by $kv_s<E_k$ shown in
Fig.~\ref{figyw1}) is still non-viscous. 

For the stationary magnetic response, when $v_s<v_L$, only
superfluid is present. In this situation, we prove that the excited superfluid
current is the Meissner supercurrent, and near the critical temperature,
our gap equation reduces to the Ginzburg-Landau equation.\cite{Ginzburg} As for
$v_s{\ge}v_L$, there exist normal fluid (from U region), non-viscous (from
P$_{\rm nv}$ region) and viscous (from P$_{\rm v}$ region) superfluids. The
magnetic response is captured by the three-fluid model proposed above. 
Specifically, differing from the excited Meissner
supercurrent in the superfluid, no current is directly excited from the magnetic
flux in the normal fluid as it should be. Nevertheless, the normal-fluid
current can be induced through the above mentioned friction drag with superfluid
current. Moreover, due to this friction, the superfluid current is separated
into the non-viscous and viscous ones. Consequently, thanks to the viscosity in
superfluid current and presence of the normal fluid current, the penetration
depth is influenced by the scattering. By only considering the viscous
superfluid, the  dependence of penetration depth on mean free path from our
theory is exactly same as that from Tinkham's discussion.\cite{Ba1}
Nevertheless, since there also exist normal fluid and non-viscous
superfluid, an extension of penetration
depth is revealed. In addition, at $v_s{\ge}v_L$, we also propose a modified
Ginzburg-Landau equation, in which the calculation of the phenomenological
parameters are restricted to the pairing region. Finally, 
at $v_s>\omega_D/k_F$ ($\omega_D$ denotes the Debye frequency) before the
superconducting gap is destroyed, we predict an exotic phase in which
the non-viscous superfluid vanishes, leaving only the viscous superfluid and
normal fluid. Thus, interestingly, this phase shows the
finite resistivity but with a {\em finite} superconducting gap.

As for the optical response, the excited superconducting velocity $v_s$ 
oscillates with time. When $v_s<v_L$, only the non-viscous superfluid is present whereas
at $v_s{\ge}v_L$, there exist normal fluid (from U region), non-viscous (from
P$_{\rm nv}$ region) and viscous (from P$_{\rm v}$ region) superfluids. 
We show that in the optical response, the normal-fluid current
exhibits the Drude-model behavior as it should be. Whereas in the superfluid,
we find that the superfluid current is excited and it consists of the Meissner
supercurrent, which 
has the same form as that in the magnetic response, as well as
the Bogoliubov quasiparticle current. At low temperature, few Bogoliubov
quasiparticles are excited in the pairing region and hence the Bogoliubov
quasiparticle current is marginal.  In this case, the normal-fluid current and
the superfluid current which only consists of Meissner supercurrent 
are exactly same as those in the original two-fluid
model.\cite{Ba0,Ba1,Ba4,L6,L7,L9,NL1,NL2,two}   
However, there exists friction between the superfluid and
normal-fluid currents. Due to this friction, the superfluid is separated
into the non-viscous and viscous ones. This suggests that the optical
response is also captured by the three-fluid model above. Then, based on this
three-fluid model, an expression of the optical conductivity is
revealed. Furthermore, we also give the expression of the optical excitation of
the Higgs mode. Comparison between the contributions from the drive and
Anderson-pseudospin pump effects mentioned above is addressed. We point out
that the previous conclusion by Yu and Wu\cite{GOBE1} that the drive effect is 
dominant only holds at finite temperature, whereas at zero temperature, both
effects contribute.    

This paper is organized as follows. In Sec.~II, we introduce our model and
construct the gauge-invariant kinetic theory of the electromagnetic response in
superconducting states. We derive the three-fluid model and perform the
analytical analysis of the magnetic and optical responses in Sec.~III. We
summarize and discuss in Sec.~IV.

\section{MODEL}
\label{model}

In this section, we first set up the Hamiltonian for the conventional
superconducting states and present the gauge structure revealed by
Nambu.\cite{gi0,gi2} Then, we extend the previous theory by Yu and
Wu,\cite{GOBE1} and present a gauge-invariant microscopic kinetic equation of
the electromagnetic response in superconducting states.

\subsection{Hamiltonian}

The free Bogoliubov-de Gennes (BdG) Hamiltonian of the $s$-wave superconducting
state reads:    
\begin{equation}
\label{BdG}
H={\int}\frac{d{\bf r}}{2}\Psi^{\dagger}(x)\{[\xi_{{{\bf p}}-e{\bf
    A}(x)\tau_3}+e\phi(x)]\tau_3+{\hat \Delta}(x)\}\Psi(x), 
\end{equation}
with 
\begin{equation}
{\hat \Delta}(x)=|\Delta|[e^{i\psi(x)}\tau_++e^{-i\psi(x)}\tau_-].
\end{equation}
Here, the Nambu-space field operator reads
$\Psi(x)=\Big(\Psi_{\uparrow}(x),\Psi^{\dagger}_{\downarrow}(x)\Big)^T$; 
$\xi_{{\bf p}}=\varepsilon_{{\bf p}}-\mu$ and $\varepsilon_{{\bf
    p}}=\frac{{\bf p}^2}{2m}$ with $m$ and $\mu$  
being the effective mass and chemical potential; ${{\bf p}}=-i\hbar{\bm {\nabla}}$;
$\tau_i$ are the Pauli matrices in particle-hole spaces. 
In the present work, we consider a magnetic flux in the magnetic response
of superconductors, and hence, the Zeeman effect of the magnetic field is neglected.

It is first revealed by Nambu that under a gauge transformation
$\Psi(x){\rightarrow}e^{i\tau_3\chi(x)}\Psi(x)$, to restore the gauge invariance of
the BdG Hamiltonian [Eq.~(\ref{BdG})], the vector 
potential ${\bf A}$, scalar potential $\phi$, and superconducting phase $\psi$
must transform as\cite{gi0,gi2}  
\begin{eqnarray}
eA_{\mu}&\rightarrow&eA_{\mu}-\partial_{\mu}\chi(x), \label{gaugestructure1}\\
\label{gaugestructure2}
\psi(x)&\rightarrow&\psi(x)+2\chi(x),
\end{eqnarray}
where the four vectors are $A_{\mu}=(\phi,{\bf A})$ and
$\partial_{\mu}=(\partial_t,-{\bm \nabla})$.

\subsection{Kinetic equation}
\label{KNE}

Following the previous work by Yu and Wu,\cite{GOBE1}  we derive the
gauge-invariant microscopic kinetic equation of the electromagnetic response
in superconducting states in the presence of the electron-electron, electron-phonon and
electron-impurity scatterings.

\subsubsection{Derivation of free Kinetic equation}
\label{DFKE}
We first present the derivation of the free kinetic
equation in the absence of the electron-electron, electron-phonon and
electron-impurity interactions.

We begin with the lesser $\tau_0$-Green function
{\small{$G^{<}_{x_1x_2}=i\langle\Psi^{\dagger}(x_2)\Psi(x_1)\rangle$}}.\cite{GOBE1}
The Gorkov's equations of the lesser
$\tau_0$-Green function $G^{<}_{x_1x_2}$ read:\cite{G1,GQ2,FF7} 
\begin{equation}
 (i\overrightarrow{\partial}_{t_1}-\overrightarrow{H}_{{{\bf p}}_1,x_1})G^{<}_{x_1x_2}=0,\label{GO1}
\end{equation}
\begin{equation}
G^{<}_{x_1x_2}(-i\overleftarrow{\partial}_{t_2}-\overleftarrow{H}_{{{\bf p}}_2,x_2})=0. \label{GO2}
\end{equation}
The gauge structure of the lesser $\tau_0$-Green function is given by 
{\small
  $G^{<}_{x_1x_2}\rightarrow{e^{i\tau_3\chi(x_1)}}G^{<}_{x_1x_2}e^{-i\tau_3\chi(x_2)}$}
after a gauge transformation
{\small{$\Psi(x){\rightarrow}e^{i\tau_3\chi(x)}\Psi(x)$}}. 
As in the kinetic equation, only the center-of-mass coordinate {\small{$R=(T,{\bf
  R})=(x_1+x_2)/2$}}  is retained.
It is hard to retain the gauge invariance in the kinetic equation derived from
$G^{<}_{x_1x_2}$. To fix this, following the previous
works,\cite{EG1,EG2,GOBE1} by introducing the Wilson line,\cite{Wilson} the
gauge-invariant Green function is constructed:   
{\small{$G^{g<}_{x_1x_2}=e^{-iW^{R}_{x_1}}G^<_{x_1x_2}e^{-iW^{x_2}_{R}}$}}. Here, 
{\small{$W^y_{x}={{\rm P}\int^y_{x}dx^{\mu}eA_{\mu}\tau_3}$}} with $dx^{\mu}=(dt,-d{\bf
  r})$. ``P'' indicates that the integral is  
path dependent. Then, after the gauge transformation
{\small{$\Psi(x){\rightarrow}e^{i\tau_3\chi(x)}\Psi(x)$}},  
$G^{g<}_{x_1x_2}$ transforms as
{\small{$G^{g<}_{x_1x_2}\rightarrow{e^{i\tau_3\chi(R)}}G^{g<}_{x_1x_2}e^{-i\tau_3\chi(R)}$}},
in which only the center-of-mass coordinate is related. 

By taking the difference of Eqs.~(\ref{GO1})
and~(\ref{GO2}) and replacing $G^{<}_{x_1x_2}$ with $G^{g<}_{x_1x_2}$, one has 
\begin{eqnarray}
&&i\widetilde{\partial_T}G^{g<}_{x_1x_2}-[e\phi(x_1)\tau_3G^{g<}_{x_1x_2}-G^{g<}_{x_1x_2}e\phi(x_2)\tau_3]
\nonumber\\
&&\mbox{}-[\tau_3{\widetilde{\xi}}_{{\overrightarrow{\bf p}_1}-e{\bf
    A}(x_1)\tau_3}G^{g<}_{x_1x_2}-G^{g<}_{x_1x_2}\widetilde{\xi}_{{\overleftarrow{\bf p}_2}-e{\bf
    A}(x_2)\tau_3}\tau_3]\nonumber\\
&&\mbox{}-[e^{-2iW^R_{x_1}}{\hat \Delta}(x_1)G^{g<}_{x_1x_2}-G^{g<}_{x_1x_2}{\hat
  \Delta}(x_2)e^{-2iW^{x_2}_{R}}]=0,~~~~
 \label{F4}
\end{eqnarray}
in which {\small
  $\widetilde{X}G^{g<}_{x_1x_2}=e^{-iW^R_{x_1}}[X(e^{iW^R_{x_1}}G^{g<}_{x_1x_2}e^{iW^{x_2}_{R}})]{e^{-iW^{x_2}_{R}}}$}
and {\small
  $G^{g<}_{x_1x_2}\widetilde{X}=e^{-iW^R_{x_1}}[(e^{iW^R_{x_1}}G^{g<}_{x_1x_2}e^{iW^{x_2}_{R}})X]{e^{-iW^{x_2}_{R}}}$}. 
Then, via taking the path in the Wilson line to be the straight
line\cite{EG1,EG2,GOBE1} and defining relative coordinate {\small{$r=(t,{\bf
      r})=x_1-x_2$}}, through the gradient expansion,\cite{GQ2,GQ3} by taking equal time, i.e,
$t=0$,\cite{GQ2,GQ3,FF7,GOBE1} the gauge-invariant kinetic equation of the
density matrix {\small $\rho_{\bf
  k}({\bf R},T)=-iG^{g<}({\bf R},T,{\bf k},t=0)=-i\int{d{\bf r}}e^{-i{\bf k}\cdot{\bf r}}G^{g<}({\bf
    R},T,{\bf r},t=0)$} is obtained from Eq.~(\ref{F4}).

It is pointed out that in the previous work by Yu and Wu,\cite{GOBE1} except the
zeroth order, the higher-order gradient expansion on the last term on the
left-hand side of Eq.~(\ref{F4}), i.e., the superconducting order 
parameter ${\hat \Delta}$ accompanied with the Wilson
line, is neglected by considering a fixed order parameter in semiconductor
quantum wells from the proximity
effect. This approximation is sublated in our work, considering the fluctuation of
$W$ and ${\Delta}$ in time and space in the
electromagnetic response.  To apply the higher-order gradient expansion on this
term, we approximately take
{\small $e^{-2iW}\approx{1-2iW-2W^2}$}. 
This approximation is based on the fact that in conventional superconductors, the
vector potential is much smaller than the Fermi momentum. 
Therefore, since one has $W\propto({\bf A}\cdot{\bf r})$ after taking equal
time, $W$ can be treated as small quantity.  

Finally, the new gauge-invariant microscopic kinetic equation of the
electromagnetic response in the superconducting states is written as
\begin{widetext}
\begin{eqnarray}
&&\partial_T\rho_{\bf
  k}+i\left[\left(\xi_k+e\phi\right)\tau_3+{\hat
  \Delta}({\bf R}),\rho_{\bf
k}\right]+i\left[\frac{e^2A^2}{2m}\tau_3,\rho_{\bf
k}\right]+\frac{1}{2}\left\{e{\bf E}\tau_3,\partial_{\bf k}\rho_{\bf
k}\right\}+\left\{\frac{\bf k}{2m}\tau_3,{\bm \nabla}_{\bf R}\rho_{\bf
    k}\right\}-\left[\frac{i}{8m}\tau_3,{\bm \nabla}^2_{\bf R}\rho_{\bf
    k}\right]\nonumber\\
&&\mbox{}-\frac{1}{2}\left\{({\bm \nabla}-2ie{\bf A}\tau_3){\hat
  \Delta}({\bf R}),\partial_{\bf k}\rho_{\bf
k}\right\}-\frac{i}{8}\left[({\bm \nabla}-2ie{\bf A}\tau_3)({\bm \nabla}-2ie{\bf A}\tau_3){\hat
  \Delta}({\bf R}),\partial_{\bf k}\partial_{\bf k}\rho_{\bf
k}\right]-\left[\frac{e{\bf
    A}}{2m}\tau_3,\tau_3{\bm \nabla}_{\bf R}\rho_{\bf
    k}\right]\nonumber\\
&&\mbox{}-\left[\frac{e{\bm
      \nabla}_{\bf R}\cdot{\bf
    A}}{4m}\tau_3,\tau_3\rho_{\bf k}\right]=\partial_t\rho_{\bf k}\Big|_{\rm sc}.\label{KE}
\end{eqnarray}\\
\end{widetext}
Here, $[A,B]=AB-BA$ and $\{A,B\}=AB+BA$ represent the commutator and
anti-commutator, respectively; ${\bf E}=-{\bm \nabla}_{\bf R}\phi-\partial_T{\bf
  A}$ denotes the electric field. It is noted that on the right-hand side of
Eq.~(\ref{KE}), the scattering term $\partial_t\rho_{\bf k}\Big|_{\rm sc}$ is
added for completeness, whose explicit expression is given in the next section.

In Eq.~(\ref{KE}), on the left-hand side, the second term represents the
coherent term contributed by the BCS Hamiltonian. The third and fourth terms denote
the pump and drive effect mentioned in the introduction, as addressed in the
previous work by Yu and Wu.\cite{GOBE1} The fifth and sixth terms stand for the
diffusion terms. The seventh and eighth terms, which behave like the drive
effect, are absent in Ref.~\onlinecite{GOBE1}. They come from the higher-order
gradient expansion of the superconducting order parameter accompanied with the
Wilson line mentioned above. In the following section, it is shown that  
these two terms provide the kinetic-energy
terms in the Ginzburg-Landau equation. Particularly, it is noted that with the
gauge structure revealed by Nambu [Eqs.~(\ref{gaugestructure1})
and~(\ref{gaugestructure2})],\cite{gi0} Eq.~(\ref{KE}) is gauge invariant after
the gauge transformation $\rho_{\bf
  k}(R)\rightarrow{e^{i\tau_3\chi(R)}}\rho_{\bf k}(R)e^{-i\tau_3\chi(R)}$.

The order parameter is self-consistently determined
by the gap equation:
\begin{equation}
\label{ge}
\Delta({\bf R})=-V{\sum_{\bf k}}'{\rm Tr}[\rho_{{\bf k}}({\bf R})\tau_-],
\end{equation}
where $V$ is the conventional $s$-wave attractive potential. ${\sum_{\bf k}}'$
here and in the following shows the summation is restricted in the
spherical shell by the BCS theory.\cite{BCS}

The gauge invariant current is obtained by performing the Wilson
line\cite{Wilson} technique on the current\cite{G1,G2} 
\begin{equation}
{\bf j}=-\frac{ie}{2m}{\rm Tr}\left[(i{\bm \nabla}_{x'}-i{\bm
  \nabla}_{x})G^<_{x,x'}-2e{\bf A}\tau_3G^<_{x,x'}\right]_{x'\rightarrow{x+0^+}},\label{C1}
\end{equation}
and reads
\begin{equation}
{\bf j}=-\frac{ie}{2m}{\rm Tr}\left[{-2i{\bm \partial}_{\bf
      r}}G^{g<}_{x,x'}\right]_{x'\rightarrow{x+0^+}}=\sum_{\bf k}{\rm
  Tr}\left[\frac{e{\bf k}}{m}\rho_{\bf k}\right].~~~\label{current}
\end{equation}

\subsubsection{Derivation of scattering}

We next present the scattering terms $\partial_t\rho_{\bf k}\Big|_{\rm sc}$ in
Eq.~(\ref{KE}) due to the electron-electron Coulomb, electron-phonon and
electron-impurity scatterings. The scattering terms are derived based on the  
generalized Kadanoff-Baym (GKB) ansatz.\cite{FF7,GQ2,GQ3,GKB} 

The specific scattering terms of the electron-electron Coulomb,
electron-phonon and electron-impurity interactions are written as (the detailed
derivation of the scattering terms can be found in the previous
works\cite{GQ3,GOBE1}) 
\begin{equation}
\partial_t\rho_{\bf k}\Big|_{\rm sc}=-\pi\sum_{\bf
  k'}\sum_{\eta_1\eta_2}[{S^{\eta_1\eta_2}_{\bf kk'}(>,<)-S^{\eta_1\eta_2}_{\bf kk'}(<,>)+{\rm
    H.c.}}],
\end{equation}
with
\begin{eqnarray}
&&S^{\eta_1\eta_2}_{\bf kk'}\Big|_{\rm ei}=n_i|V_{\bf kk'}|^2\delta(E^{\eta_1}_{{\bf k}'}-E^{\eta_2}_{{\bf
    k}})\left[\tau_3\rho^>_{{\bf k}'}\Gamma^{\eta_1}_{{\bf k}'}\tau_3\Gamma^{\eta_2}_{{\bf k}}\rho^<_{\bf
k}\right],\nonumber\\
&& \\
&&S^{\eta_1\eta_2}_{\bf kk'}\Big|_{\rm ep}=|g^{\gamma_p}_{\bf kk'}|^2\Big[n^>_{\bf k-k'}\delta(E^{\eta_1}_{{\bf k}'}-E^{\eta_2}_{{\bf
    k}}+\omega^{\gamma_p}_{\bf k-k'})+n^<_{\bf k-k'}\nonumber\\
&&\mbox{}\times\delta(E^{\eta_1}_{{\bf k}'}-E^{\eta_2}_{{\bf
    k}}-\omega^{\gamma_p}_{\bf k-k'})\Big]\left[\tau_3\rho^>_{{\bf k}'}\Gamma^{\eta_1}_{{\bf k}'}\tau_3\Gamma^{\eta_2}_{{\bf k}}\rho^<_{\bf
k}\right],\\
&&S^{\eta_1\eta_2}_{\bf kk'}\Big|_{\rm ee}=\sum_{\bf q}\sum_{\eta_3\eta_4}|V_{\bf q}|^2\delta(E^{\eta_1}_{\bf
  k-q}-E^{\eta_2}_{\bf k}+E^{\eta_3}_{{\bf k}'+{\bf q}}-E^{\eta_4}_{{\bf k}'}) \nonumber\\
&&\mbox{}\times\left[\tau_3\rho^>_{\bf k-q}\Gamma^{\eta_1}_{{\bf k-q}}\tau_3\Gamma^{\eta_2}_{{\bf k}}\rho^<_{\bf
k}\right]{\rm Tr}\left[\rho^>_{{\bf k}'+{\bf q}}\Gamma^{\eta_3}_{{\bf k}'+{\bf q}}\Gamma^{\eta_4}_{{\bf k}'}\rho^<_{\bf
k'}\right].
\end{eqnarray}
Here,  $\eta=\pm$; $\Gamma^{\pm}_{\bf k}$ represent
the projection operators; $n_i$ is the impurity density; $V_{\bf q}$ denotes the
screened Coulomb 
potential; $g^{\gamma_p}_{\bf kk'}$ stands for the electron-phonon interaction
and $\omega^{\gamma_p}_{\bf q}$ represents the phonon energy 
with $\gamma_p$ being the corresponding phonon branch; $\rho^<_{\bf k}=\rho_{\bf
  k}$ and $\rho^>_{\bf k}=1-\rho_{\bf k}$; $n^>_{\bf k}=1+n_{\bf k}$
and $n^<_{\bf k}=n_{\bf k}$ with $n_{\bf k}$ being the phonon
distribution function.

As mentioned in the introduction, it is
established\cite{FF4,FF5,FF6,FF7,FF8,FF9} that with the
superconducting velocity ${\bf v}_s$,
the quasiparticle energy is tilted as {\small $E^{\pm}_{\bf
    k}={\bf k}\cdot{\bf v}_s\pm{E_k}$} with {\small $E_{k}=\sqrt{\xi^2_k+|\Delta|^2}$}.
In this situation, the projection operators are written as {\small $\Gamma^{\pm}_{\bf
    k}=U^{\dagger}_kQ^{\pm}U_k$} with {\small
  $Q^{\pm}=(1\pm\tau_3)/{2}$}. $U_k=u_k\tau_0-v_k\tau_++v_k\tau_-$  
represents the unitary transformation matrix from the particle space to the
quasiparticle one with $u_k=\sqrt{1/2+\xi_k/(2E_k)}$ and
$v_k=\sqrt{1/2-\xi_k/(2E_k)}$. It is noted that the effect of the
superconducting velocity on the scattering process is neglected in
Ref.~\onlinecite{GOBE1} by taking the quasiparticle energies as the BCS ones (i.e.,
$E^{\pm}_{\bf k}={\pm}E_k$).

\section{ANALYTICAL ANALYSIS}

In this part, with the new gauge-invariant microscopic kinetic equation
[Eq.~(\ref{KE})] in Sec.~\ref{KNE}, we analytically investigate the
electromagnetic properties of superconductors including the
magnetic and optical responses in the linear and nonlinear regimes in the
weak scattering limit.

\subsection{Weak scattering}

We first simplify the scattering terms by transforming the scattering terms into
the quasiparticle space (i.e., $\partial_t\rho_{\bf k}\Big|_{\rm sc}=U_{k}\partial_t\rho^q_{\bf
  k}\Big|_{\rm sc}U^{\dagger}_{k}$). Considering the fact that the
electron-phonon scattering is weak at low temperature, we mainly consider the
electron-impurity scattering, which reads:
\begin{eqnarray}
\partial_t\rho^q_{\bf k}\Big|_{\rm sc}&=&-n_i\pi\sum_{\bf
  k'}|V_{\bf kk'}|^2U^{\dagger}_{k}\tau_3U_{k'}\Big\{Y_{kk'}(\rho^{q}_{\bf k}-\rho^q_{{\bf k}'})\nonumber\\
&&-[\rho^q_{\bf
  k'},Y_{kk'}]\Big\}+{\rm H.c.}.\label{s1}
\end{eqnarray}
Here, $Y_{kk'}=\sum_{\eta_1\eta_2}Q^{\eta_1}U^{\dagger}_{k'}\tau_3U_{k}Q^{\eta_2}\delta(E^{\eta_1}_{{\bf k}'}-E^{\eta_2}_{{\bf
    k}})$. 

In the present work, we consider a weak scattering limit. In this situation, the
scattering only causes the momentum (current) relaxation. Therefore, one only 
needs to keep the leading contribution in the scattering terms, i.e.,  the
diagonal terms in $\rho^q_{\bf k}$ (quasiparticle distribution) and
$\partial_t\rho^q_{\bf k}\Big|_{\rm sc}$ (scattering of the quasiparticle
distribution), and Eq.~(\ref{s1}) becomes
\begin{widetext}
\begin{eqnarray}
&&\partial_t\rho^q_{\bf k}\Big|_{\rm sc}=-n_i\pi\sum_{\bf
  k'}|V_{\bf kk'}|^2\Bigg\{(1-\eta_{kk'})\left(\begin{array}{cc}
(\rho^{q}_{{\bf k},11}-\rho^q_{{\bf k}',11})\delta(E^{+}_{{\bf k}'}-E^{+}_{{\bf
    k}}) & 0\\
0& (\rho^{q}_{{\bf k},22}-\rho^q_{{\bf k}',22})\delta(E^{-}_{{\bf k}'}-E^{-}_{{\bf
    k}})
\end{array}\right)\nonumber\\
&&\mbox{}+(1+\eta_{kk'})\left(\begin{array}{cc}
(\rho^{q}_{{\bf k},11}-\rho^q_{{\bf k}',22})\delta(E^{-}_{{\bf k}'}-E^{+}_{{\bf
    k}}) & 0\\
0& (\rho^{q}_{{\bf k},22}-\rho^q_{{\bf k}',11})\delta(E^{+}_{{\bf k}'}-E^{-}_{{\bf
    k}})
\end{array}\right)\Bigg\} ,\label{s2}
\end{eqnarray}\\
\end{widetext}
where $\eta_{kk'}=(|\Delta|^2-\xi_k\xi_{k'})/(E_kE_{k'})$.

\begin{figure}[htb]
  {\includegraphics[width=8.0cm]{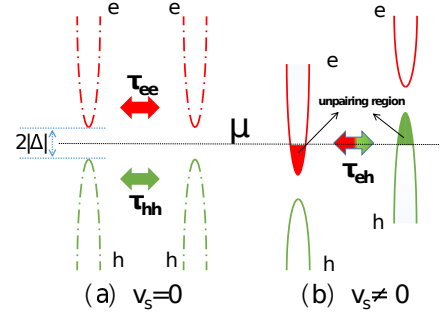}}
\caption{(Color online) Schematic showing the tilt of the quasiparticle energy spectrum
  and scattering processes. The chain (solid) curves represent the
  quasiparticle energy in the absence (presence) of a large
  superconducting velocity. The filled arrows represent the scattering process.
  In (a), the inter-band scattering between the quasielectrons and quasiholes
  is forbidden by the energy conservation. Only the intra quasielectron-band (denoted by $\tau_{ee}$) and
  quasihole-band (denoted by $\tau_{hh}$) scatterings
  exist. In (b), the presence of the large superconducting velocity ($kv_s>E_k$) tilts the
  quasiparticle energy spectrum and hence the unpairing regions (represented by
  red and green regions) emerge. In this case, 
  the inter-band scattering between the quasielectrons and quasiholes (denoted
  by $\tau_{eh}$) is turned on. } 
  
\label{figyw2}
\end{figure}

On the right-hand side
of Eq.~(\ref{s2}), the first term denotes the intra quasielectron-band and
intra quasihole-band scatterings. The second term represents the inter-band
scattering between the quasielectrons and quasiholes. Actually, as shown in
Fig.~\ref{figyw2} (a), in the absence of the superconducting velocity, the
inter-band scattering between the quasielectrons and quasiholes is forbidden by
the energy conservation thanks to the BCS gap. Only the intra-band scatterings
exist. Nevertheless, as mentioned above, with a large excited superconducting
velocity ($kv_s>E_k$) in the electromagnetic response,\cite{GOBE1,GOBE3} the
quasiparticle energy spectrum is tilted.\cite{FF4,FF5,FF6,FF7,FF8,FF9} Then, as
shown in Fig.~\ref{figyw2} (b), the inter-band scattering between the
quasielectrons and quasiholes is turned on. However, this unique scattering has
long been overlooked in the literature.

In conventional superconducting metals, due to the strong screening, 
one can take the impurity scattering as the short-range one,
i.e., $|V_{\bf kk'}|^2\approx|V_0|^2$. Moreover, thanks to the large Fermi
energy, we approximately take the emergence of the scattering around the Fermi 
surface by setting $|\xi_{k}|,|\xi_{k'}|<E_c$ in Eq.~(\ref{s2}). $E_c$ is the
cutoff energy. Then, after the integration over the angle, Eq.~(\ref{s2})
approximately becomes (refer to Appendix~\ref{aa}) 
\begin{eqnarray}
\partial_t\rho^q_{\bf k}\Big|_{\rm
  scat}&=&-\frac{1}{\tau_k}\Bigg[\frac{1+\tau_3}{2}(\rho^{q}_{{\bf k},11}-\rho^q_{{\bf
    k}',22})\Big|^{|\xi_k|=|\xi_{k'}|}_{\delta\theta_{\bf kk'}=\frac{E_k}{kv_s}}\nonumber\\
&&\mbox{}+\frac{1-\tau_3}{2}(\rho^{q}_{{\bf k},22}-\rho^q_{{\bf
    k}',11})\Big|^{|\xi_k|=|\xi_{k'}|}_{\delta\theta_{\bf
    kk'}=-\frac{E_k}{kv_s}}\Bigg].~~~\label{s5} 
\end{eqnarray}
Here,
$1/\tau_k=2n_i\pi|V_0|^2D_0\lambda_{c}(1+4u^2_kv^2_k)$ with $D_0=mk_F/(2\pi^2)$ 
denoting the density of states and $\lambda_c$ being a dimensionless parameter; 
$\delta\theta_{\bf kk'}=(\cos\theta_{{\bf k}'}-\cos\theta_{\bf
  k})/2$. Consequently, the scattering term is simplified.

\subsection{Three-fluid model}
\label{tfm}
Based on Eq.~(\ref{s5}), we next perform an analysis on the scattering and
derive a three-fluid model in the electromagnetic response in the
superconducting states. Specifically, it is noted that from Eq.~(\ref{s5}), one
always has 
$|\delta\theta_{\bf kk'}|={E_k}/({kv_s})$. Therefore, since
$|\delta\theta_{\bf kk'}|=|(\cos\theta_{{\bf k}'}-\cos\theta_{\bf k})|/2{\le}1$,  
the scattering term is nonzero only in the region $kv_s>E_k$. 
This is natural since when $kv_s>E_k$, as mentioned in the introduction, 
unpairing (U) region with $|{\bf k}\cdot{\bf v}_s|>E_k$, in which the particles
no longer participate in the pairing and behave like the normal particles,
emerges.\cite{FF7,FF8,FF9,FF1} Then, the normal fluid is present. Hence, 
the scattering in the unpairing (U) region is nonzero. Consequently, a threshold
of superconducting velocity $v_s$  for the emergence of
normal fluid and hence scattering is predicted from our theory as
\begin{equation}
v_L=\frac{|\Delta|}{k_F}.
\end{equation}
As mentioned in the introduction, we refer to this threshold in superconducting
state as Landau threshold, following Landau in bosonic liquid helium II theory.\cite{Landau}   

Besides U region, there also exists special pairing region (P$_{\rm v}$ region)
with $kv_s>E_k$ and $|{\bf k}\cdot{{\bf v}_s}|<E_k$, in which the scattering is
also finite since $kv_s>E_k$. This is due to the fact that this region share the 
same momentum magnitude with U region, as shown in Fig.~\ref{figyw1}. Since the
short-range impurity scattering 
is isotropic in the momentum space, the particles in P$_{\rm v}$
region participate in the pairing but experience the scattering with those in U region, and hence the
superfluid from P$_{\rm v}$ region becomes viscous. This can also be
understood as follows. In the first term on the right-hand side 
of Eq.~(\ref{s5}), the particle with ${\bf k}$ is scattered by that with ${\bf
  k}'$. When the ${\bf k}$ particle is in P$_{\rm v}$ region
($|{\bf k}\cdot{{\bf v}_s}|<E_{k}$ but $kv_s>E_{k}$), one has $-3E_{k'}<{\bf
 k}'\cdot{{\bf v}_s}<E_{k'}$, and hence, the ${\bf k}'$ particle sits in
U region. This indicates that the particles in P$_{\rm v}$ region experience the
scattering from those in U region. By using similar analysis, one can find that 
the particles in U region experience the scattering from those in {\em both} U and
P$_{\rm v}$ regions. The internal scattering in U region is natural since the
particles in U region behave like the normal ones. Whereas the inter scattering 
between P$_{\rm v}$ and U regions denotes the existence of the friction between the
superfluid and normal fluid. Therefore, the superfluid from P$_{\rm v}$ region
becomes viscous. As for the remaining pairing region (P$_{\rm nv}$ region with
$kv_s<E_k$), the superfluid in this region is still non-viscous. 

Consequently, a three-fluid model for the electromagnetic response in the
superconducting states at $v_s{\ge}v_L$ is predicted from our theory: normal
fluid (from U region), non-viscous (from P$_{\rm nv}$ region) and viscous (from
P$_{\rm v}$ region) superfluids. Based on this three-fluid model, in the
following sections, we show that the electromagnetic properties of the
superconducting states including both the magnetic and optical
responses can be well captured.

\subsection{Magnetic response}

In this part, by using the gauge-invariant kinetic equation, we investigate the 
stationary magnetic response in the superconducting states. Properties of
the excited current and superconducting order parameter are addressed.  

\subsubsection{Solution of density matrix}

In the stationary situation, one has $\partial_t\rho_{\bf k}=0$, $\phi=0$ and $e{\bf E}=0$
in kinetic equation. By expanding the density matrix as $\rho_{\bf
  k}=\rho_{{\bf k}0}\tau_0+\rho_{{\bf k}-}\tau_-+\rho_{{\bf k}+}\tau_++\rho_{{\bf k}3}\tau_3$, Eq.~(\ref{KE}) becomes 
\begin{eqnarray}
&&\left(\varepsilon_{k}-\mu+\frac{\varepsilon_{{\bf p}-2e{\bf
        A}}}{4}\right)\rho_{{\bf k}+}=\bigg[\rho_{{\bf k}3}-\frac{i{\bm \partial}_{\bf k}\rho_{{\bf
    k}0}\cdot({\bm \nabla}-2ie{\bf A})}{2}\nonumber\\
&&\mbox{}-\frac{{\bm \partial}_{\bf k}{\bm \partial}_{\bf k}\rho_{{\bf
    k}3}:({\bm \nabla}-2ie{\bf A})({\bm \nabla}-2ie{\bf
  A})}{8}\bigg]\Delta+\frac{\left\{\partial_{t}\rho_{\bf k}\Big|_{\rm sc}\right\}_{+}}{2},\nonumber\\
&&\mbox{} \label{E7} \\
&& \frac{\bf k}{m}\cdot{\bm \nabla}\rho_{{\bf k}0}=i\Delta^*\rho_{{\bf
    k}+}-i\Delta\rho_{{\bf k}-}+\left\{\partial_{t}\rho_{\bf k}\Big|_{\rm
  sc}\right\}_{3}\nonumber\\
&&\mbox{}+\frac{i{\bm \partial}_{\bf k}{\bm \partial}_{\bf k}\rho_{{\bf
    k}-}:({\bm \nabla}-2ie{\bf A})({\bm \nabla}-2ie{\bf
  A})\Delta}{8}\nonumber\\
&&\mbox{}-\frac{i{\bm \partial}_{\bf k}{\bm \partial}_{\bf k}\rho_{{\bf
    k}+}:({\bm \nabla}+2ie{\bf A})({\bm \nabla}+2ie{\bf
  A})\Delta^*}{8},\label{E8}
\end{eqnarray}
with $\rho_{{\bf k}-}=\rho^*_{{\bf k}+}$. Since $\mu{\gg}\varepsilon_{{\bf p}-2e{\bf
    A}}$ thanks to the large Fermi energy in the conventional superconductors,
$\varepsilon_{{\bf p}-2e{\bf A}}$ on the left-hand side of Eq.~(\ref{E7}) can be neglected. 

Then, from
Eqs.~(\ref{E7}) and~(\ref{E8}), by only keeping the diagonal terms in the
density matrix in the quasiparticle space due to their leading contribution, the
solution of the density matrix in 
the quasiparticle space is obtained as (refer to Appendix~\ref{ab})
\begin{eqnarray}
\rho^q_{\bf k}&=&\left(\begin{array}{cc}
f(E^+_{\bf k})& 0\\
0 &f(E^-_{\bf k}) 
\end{array}\right)+({\bf k}{\cdot}{\bf v}_s)\left(\begin{array}{cc}
a^+_{\bf k}& 0 \\
0 &a^-_{\bf k} 
\end{array}\right)\nonumber\\
&&+({\bf k}{\cdot}{\bf v}_s)|\Delta|^2\left(\begin{array}{cc}
m^+_{\bf k}& 0\\
0 &m^-_{\bf k}
\end{array}\right)+\frac{({\bf k}{\cdot}{\bf v}_s)^2}{2}\left(\begin{array}{cc}
b^+_{\bf k}& 0\\
0 &b^-_{\bf k} 
\end{array}\right)\nonumber\\
&&+({\bf k}{\cdot}{\bf v}_s)|\Delta|^2\left(\begin{array}{cc}
{\delta}m^+_{\bf k}& 0\\
0 &{\delta}m^-_{\bf k} 
\end{array}\right), \label{mr}
\end{eqnarray}
with 
\begin{eqnarray}
&&a^{\pm}_{\bf k}=\mp{\partial}_{E_k}f(E^{\pm}_{\bf k}),\\
&&b^{\pm}_{\bf
  k}={\partial}^2_{E_k}f(E^{\pm}_{\bf k})+\frac{{\partial}_{E_k}f(E^{\pm}_{\bf
  k})}{E_k},\\
&&m^{\pm}_{\bf
    k}=\pm\left[\frac{1}{E_k}\partial_{E_k}+\frac{1}{4\xi_k\varepsilon_k}\right]\frac{f(E^{\pm}_{\bf
    k})}{E_k},\\
&&{\delta}m^{\pm}_{\bf
  k}=\mp\frac{\xi}{\tau_kv_F}\theta\left(\frac{kv_s}{E_k}\right)\frac{\xi_k}{E_k}m^{\pm}_{\bf
    k}
\end{eqnarray}
Here, ${\bf v}_s={\bf p}_s/{m}$ (refer to Appendix~\ref{ab}); the
gauge invariant ${\bf
  p}_s={\bm \nabla}\psi/2-e{\bf A}$ denotes the superconducting
momentum;\cite{gi0,gi2,GOBE1,GOBE2,GOBE3} $f(x)$
represents the Fermi distribution; $\theta(x)$ is the step function. 

As seen from Eq.~(\ref{mr}), the first term in $\rho^q_{\bf k}$ represents the
quasiparticle distribution of the FFLO-like state. The second term stands for the
linear response of the quasiparticle state. The third term denotes the
Meissner-supercurrent response, which is proved in the following. The forth term
represents the nonlinear response. The last term is the scattering
contribution, which emerges at $kv_s>E_k$ as mentioned in Sec.~\ref{tfm}.

\subsubsection{Excited current}
\label{ECIM}
With Eqs.~(\ref{mr}) and (\ref{current}), by neglecting the nonlinear response,
the excited current in the stationary magnetic response reads:  
\begin{eqnarray}
{\bf j}&=&\frac{2e}{m}\sum_{\bf k}{\bf k}\rho_{{\bf k}0}=\frac{2e}{m}\sum_{\bf k}{\bf k}\rho^q_{{\bf k}0}\nonumber\\
&=&\frac{2e}{m}\sum_{\bf k}{\bf k}\Bigg[\frac{f(E^+_{\bf k})+f(E^-_{\bf
    k})+({\bf k}{\cdot}{\bf v}_s)(a^+_{\bf k}+a^-_{\bf k})}{2}\nonumber\\
&&\mbox{}+({\bf k}{\cdot}{\bf v}_s)|\Delta|^2\frac{m^+_{\bf k}+m^-_{\bf k}+{\delta}m^+_{\bf k}+{\delta}m^-_{\bf k}}{2}\Bigg].\label{c1}
\end{eqnarray}

When $v_s<{v_L}$, no U region emerges and the momentum space belongs to
non-viscous pairing (P$_{\rm nv}$) region.  Therefore, only the non-viscous
superfluid is
present. Then,
one has {\small{$f(E^{\pm}_{\bf k})\approx{f(\pm{E_k})+({\bf k}{\cdot}{\bf
    v}_s)\partial_{E_k}f(E_k)}$}}, and Eq.~(\ref{c1}) becomes 
\begin{eqnarray}
{\bf j}&=&e{\bf v}_sD_0\int\frac{d\Omega}{4\pi}\cos^2\theta_{\bf
  k}\int{d\xi_k}\left(4\varepsilon_{k_F}|\Delta|^2\frac{m^+_{\bf k}+m^-_{\bf k}}{2}\right)\nonumber\\
&=&e{\bf v}_sD_0\int\frac{d\Omega}{4\pi}\cos^2\theta_{\bf
  k}\int{d\xi_k}\rho_{m{\bf k}},\label{c2}
\end{eqnarray}
with 
\begin{equation}
\label{rmk}
\rho_{m{\bf
    k}}=\frac{4\varepsilon_{k_F}|\Delta|^2}{E_k}\partial_{E_k}\left[\frac{f(E^+_k)-f(E^-_k)}{2E_k}\right].
\end{equation}
In the pairing region, with {\small{$\rho_{m{\bf
    k}}\approx\frac{4\varepsilon_{k_F}|\Delta|^2}{E_k}\partial_{E_k}[\frac{2f(E_k)-1}{2E_k}]$}},
the current reads: 
\begin{equation}
{\bf j}=e{\bf
    v}_sN_0|\Delta|^2\frac{7R(3)}{4(\pi{T})^2},
\end{equation}
which is exactly same as the Meissner supercurrent in the literature.\cite{Gor-G,G1} Here, $R(x)$
is Riemann zeta function and $N_0$ represents the electron
density. Consequently, we refer to $\rho_{m{\bf k}}$ as 
Meissner-superfluid density. Particularly, it is noted that the excited Meissner
supercurrent entirely comes from $m^{\pm}_{\bf k}$ terms, indicating that the third
term in Eq.~(\ref{mr}) gives rise to the Meissner-supercurrent response.   

For the case $v_s>v_L$, as mentioned in Sec.~\ref{tfm}, there exist the normal
fluid (from U region), non-viscous (from P$_{\rm nv}$ region) and viscous (from
P$_{\rm v}$ region) superfluids. In this situation, considering
the fact that {\small{$f(E^{\pm}_{\bf k})\approx{f(\pm{E_k})+({\bf k}{\cdot}{\bf 
    v}_s)\partial_{E_k}f(E_k)}$}} in P$_{\rm v}$ and P$_{\rm nv}$ regions and
{\small{$f(E^{\pm}_{\bf k})\approx{f({\bf k}\cdot{\bf v}_s)\pm{E_k\partial_{{\bf
            k}\cdot{\bf v}_s}}f({\bf k}{\cdot}{\bf  
    v}_s)}$}} in U region, with  $f({\bf k}{\cdot}{\bf  
    v}_s)\approx{f(0)}+({\bf k}{\cdot}{\bf  
    v}_s)\partial_{0}f(0)$ near the Fermi surface,  Eq.~(\ref{c1}) becomes 
\begin{equation}
{\bf j}={\bf j}_{\rm P_{nv}}+{\bf j}_{\rm P_{v}}+{\bf j}_{\rm U},\label{c3}
\end{equation}
where
\begin{eqnarray}
{\bf j}_{\rm P_{nv}}&=&e{\bf v}_s\sum_{{\bf k}\in{\rm P_{nv}}}\rho_{{m}{\bf k}}\cos^2\theta_{\bf
  k},\\
{\bf j}_{\rm P_{v}}&=&e{\bf v}_s\sum_{{\bf k}\in{\rm P_{v}}}\left(1-\frac{\xi}{l}\right)\rho_{{m}{\bf k}}\cos^2\theta_{\bf
  k},\\
{\bf j}_{\rm U}&=&-e{\bf v}_s\sum_{{\bf k}\in{\rm U}}\frac{\xi}{l}\rho_{{m}{\bf k}}\cos^2\theta_{\bf
  k}.
\end{eqnarray}
Here,
{\small{$l=3N_0{\tau_kv_F}/\pi^3[1/(2D_0E_k)+\partial_{(D_0E_k)}f(E_k)]$}}
denotes the mean-free path in the superconducting states.

The features of Eq.~(\ref{c3}) can be well captured by the three-fluid model
described in Sec.~\ref{tfm}. Specifically, without the scattering ($1/l=0$), 
the Meissner supercurrent (${\bf j}_{\rm P_{nv}}$+${\bf j}_{\rm P_{v}}$) is
excited in the superfluid (P$_{\rm v}$ and P$_{\rm nv}$ regions) whereas no
current (${\bf j}_{\rm U}=0$ when $1/l=0$) is directly excited from the magnetic
flux in the normal fluid (U region) as it should be. Nevertheless, in the
presence of the scattering ($1/l\ne0$), the normal-fluid current ${\bf j}_{\rm
  U}$ can be induced through the friction drag with the superfluid current
mentioned in Sec.~\ref{tfm}. Moreover, due to this friction, the superfluid
current ${\bf j}_{\rm P_{v}}$ becomes viscous while ${\bf j}_{\rm P_{nv}}$ is
still non-viscous.

Thanks to the normal-fluid and viscous-superfluid currents, the penetration
depth is influenced by 
the scattering. Particularly, by only considering the viscous superfluid current
${\bf j}_{\rm P_{v}}$, the penetration depth reads
$\delta^2=\delta^2_{c}/(1-\xi/l)\approx\delta^2_{c}(1+\xi/l)$ at the weak
scattering, exactly same as the one from Tinkham's discussion.\cite{Ba1}
Nevertheless, since there also exists the normal-fluid current induced by
friction drag and non-viscous superfluid, the dependence of
penetration depth becomes
\begin{equation}
\delta^2={\delta^2_{c}}{(1+\xi/l_{\rm eff})},
\end{equation}
with the clean-limit penetration depth $\delta_{c}$ and effective mean-free
path $l_{\rm eff}$ given by
\begin{equation}
{\delta_{c}}=\big(e^2\sum_{{\bf k}\in{\rm P}}\rho_{{m}{\bf k}}\cos^2\theta_{\bf
  k}\big)^{-\frac{1}{2}},
\end{equation}
\begin{equation}
\frac{1}{l_{\rm eff}}=\frac{\sum_{{\bf k}\in({\rm P_{v}}+{\rm U})}\frac{\rho_{{m}{\bf k}}\cos^2\theta_{\bf
  k}}{l}}{{\sum_{{\bf k}\in{\rm P}}\rho_{{m}{\bf k}}\cos^2\theta_{\bf
  k}}},
\end{equation}
respectively.

\subsubsection{Modified Ginzburg-Landau equation}
\label{MGLE}
In this part, we investigate the stationary magnetic response of the
superconducting order parameter. We first focus on the case at $v_s<v_L$, in
which only the non-viscous superfluid is present. In this situation, we prove that the gap
equation in our theory [Eq.~(\ref{ge})] exactly reduces to the Ginzburg-Landau 
theory\cite{Ginzburg,Gor-G,G1} (refer to Appendix~\ref{ac}). 

We next focus on the situation at $v_s>v_L$, in which both the normal fluid and
superfluid are present. Specifically, with Eq.~(\ref{mr}), from the gap equation
[Eq.~(\ref{KE})], one has
\begin{eqnarray}
\Delta&=&V{\sum_{\bf k}}'\Bigg[-\frac{\Delta}{E_k}\rho^q_{{\bf k}3}\Bigg]\nonumber\\
&=&-V{\sum_{\bf k}}'\frac{\Delta}{E_k}\Bigg[\frac{f(E^+_{\bf k})-f(E^-_{\bf
    k})}{2}+({\bf k}{\cdot}{\bf v}_s)\frac{a^+_{\bf k}-a^-_{\bf
    k}}{2}\nonumber\\
&&\mbox{}+\frac{({\bf k}{\cdot}{\bf v}_s)^2}{2}\frac{b^+_{\bf k}-b^-_{\bf
    k}}{2}+({\bf k}{\cdot}{\bf v}_s)|\Delta|^2\frac{m^+_{\bf k}-m^-_{\bf
    k}}{2}\nonumber\\
&&\mbox{}+({\bf k}{\cdot}{\bf v}_s)|\Delta|^2\frac{{\delta}m^+_{\bf k}-{\delta}m^-_{\bf
    k}}{2}\Bigg]. \label{g1}
\end{eqnarray}
By using the same expansion of $f(E^{\pm}_{\bf k})$ in each regions in Sec.~\ref{ECIM}, Eq.~(\ref{g1}) becomes 
\begin{equation}
\Delta\Bigg[\sum_{{\bf k}\in{\rm P}}\frac{1-2f(E_k)}{2E_k}-\frac{1}{V}\Bigg]-m{v^2_s\Delta}\lambda=0,\label{g2}
\end{equation}
where
\begin{equation}
  \label{kinetic}
\lambda=\varepsilon_{k_F}\Bigg[\sum_{{\bf k}\in{\rm P}}\frac{\cos^2\theta_{\bf k}\partial^2_{E_k}f(E_k)}{E_k}\Bigg].
\end{equation}
Near the critical temperature, the superconducting order parameter can be treated as
small quantity. Then, with ${\bf
  v}_s\Delta=({\nabla\psi-2e{\bf A}})\Delta/(2m)\approx(-i{\bm
  \nabla}-2e{\bf A})/(2m)$, Eq.~(\ref{g2}) can be transformed into
\begin{equation}
\label{MGL}
\Bigg\{\frac{\lambda({\bm
  \nabla}-2ie{\bf A})^2}{4m}+\left[\alpha-\beta|\Delta|^2\right]\Bigg\}\Delta=0,
\end{equation}
with 
\begin{eqnarray}
\alpha&=&\sum_{{\bf k}\in{\rm P}}\Bigg[\frac{1-2f(E_k)}{2E_k}\Bigg]\Bigg|_{|\Delta|=0}-\frac{1}{V},\\
\beta&=&\sum_{{\bf k}\in{\rm P}}\Bigg\{\frac{1}{2E_k}\partial_{E_k}\Bigg[\frac{2f(E_k)-1}{2E_k}\Bigg]\Bigg\}\Bigg|_{|\Delta|=0}.  
\end{eqnarray}
Consequently, a modified Ginzburg-Landau theory is obtained. Particularly, it
is noted that calculation of the phenomenological parameters $\alpha$ and
$\beta$ are restricted to the pairing (P) region. 

\begin{figure}[htb]
  {\includegraphics[width=7.8cm]{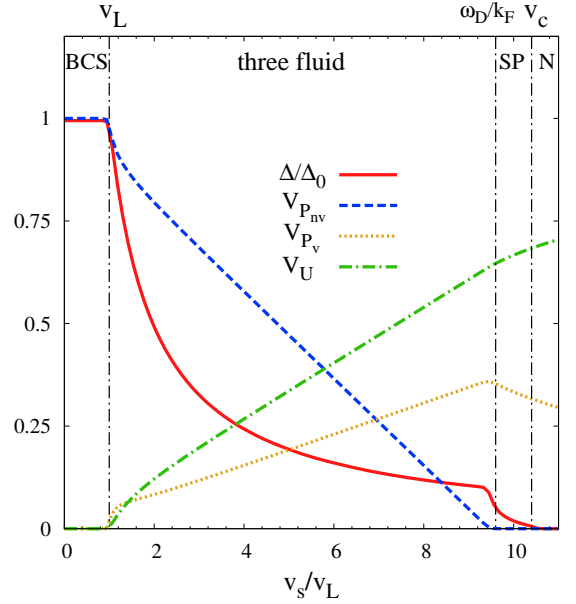}}
\caption{(Color online) Superconducting order parameter $\Delta$ and volume proportions
  of the unpairing region $V_{\rm U}$, viscous $V_{\rm P_v}$ and non-viscous
  $V_{\rm P_{nv}}$ pairing regions versus superconducting velocities $v_s$. 
  The order parameter is self-determined from the gap equation
  [Eq.~(\ref{g2})]. $\Delta_0$ denotes the BCS superconducting order parameter at
  zero temperature. The used parameter in the calculation includes $E_F=1.021~$eV,\cite{Pb1}
  $\omega_D=10.75~$meV,\cite{Pb2} $m=m_e$, $\Delta_0=1.13~$meV\cite{Pb2}
  and $T=0.02~$K. $m_e$ 
  represents the free electron mass. $v_L=|\Delta_0|/k_F$. The vertical chain
  line stands for the crossover. N denotes the normal state. SP represents the
  special phase with both finite resistivity and order parameter. $v_c$ denotes
  the critical point into the normal state. } 
  
\label{figyw3}
\end{figure}

\subsubsection{Exotic phase with both finite resistivity and order parameter}

In this part, we show the volume proportion of the unpairing region ($V_{\rm
  U}={\sum_{{\bf k}\in{\rm U}}\Xi^{-1}}$), viscous ($V_{\rm
  P_v}={\sum_{{\bf k}\in{\rm P_v}}{\Xi^{-1}}}$) and non-viscous ($V_{\rm
  P_{nv}}={\sum_{{\bf k}\in{\rm P_{nv}}}{\Xi^{-1}}}$) pairing
regions during the magnetic response in Fig.~\ref{figyw3}
by performing a numerical calculation for a specific material Pb
through self-consistently solving the gap equation [Eq.~(\ref{g2})].  
Here, $\Xi={\sum_{\bf k}}'1$ is the volume of the spherical shell.  
As seen from the figure, when $v_s<v_L$, only the non-viscous superfluid ($V_{\rm P_{nv}}\ne0$) is
present. When $v_L<v_s<9.5v_L\approx\omega_D/k_F$, the finite $V_{\rm P_{nv}}$, $V_{\rm
  P_{v}}$ and $V_{\rm U}$ indicate that there exist the normal fluid (from U
region), non-viscous (from P$_{\rm nv}$ region) and viscous (from P$_{\rm v}$
region) superfluids. Actually, in most conventional superconducting materials,
due to the large $k_F$, the value of the Landau threshold $v_L$ is very small
(for Pb, one has $v_L\approx0.33~$nm/ps at $T=0~$K and the corresponding vector
potential is $eA\approx2.9\times10^{-3}~$/nm) and hence hard to be detected.

\begin{figure}[htb]
  {\includegraphics[width=8.0cm]{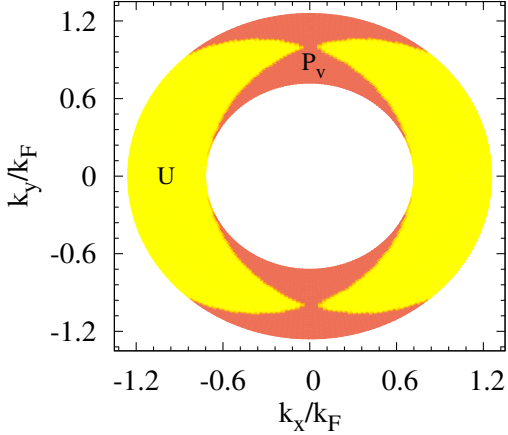}}
\caption{(Color online) Schematic showing the division in the momentum space (in
  the spherical shell by the BCS theory) when $v_{s}>\omega_D/k_F$. In
  this situation,  the spherical shell by the BCS
theory is divided into only two parts: U region, denoted
by yellow regions; P$_{\rm nv}$ region,  denoted by purple region. } 
\label{figyw4}
\end{figure}

Interestingly, before the superconducting gap $|\Delta|$ becomes zero
(i.e., at $v_s<v_c$ where $v_c$ denotes the critical point into the normal
state and $v_c\approx10.4v_L$ here from the self-consistent calculation),
with the increase of $v_s$ after $\omega_D/k_F\approx9.5v_L$, we
find that the superconducting state falls into a special phase, in which the
non-viscous superfluid vanishes ($V_{\rm P_{nv}}=0$), leaving only the viscous
superfluid ($V_{\rm P_{v}}\ne0$) and normal fluid  ($V_{\rm U}\ne0$).
This is because that the increase of $v_s$ at
$v_s>v_L$ enlarges U and hence P$_{\rm v}$ regions.
When $v_s>\omega_D/k_F$, as shown in Fig.~\ref{figyw4}, 
the spherical shell by the BCS theory is filled with U and P$_{\rm v}$
regions and P$_{\rm nv}$ region (non-viscous superfluid)
vanishes. Particularly, due to the absence of the non-viscous superfluid, the
resistivity in this phase is finite but the superconducting gap is
{\em finite}.    

In high-temperature superconductors\cite{HTSC1,HTSC2,HTSC3,HTSC4,HTSC5,HTSC6} and strongly
disordered superconductors,\cite{SDSC1,SDSC2,SDSC3,SDSC4}  the phase with both
finite resistivity and gap, known as pseudogap phase, has been widely
studied. In the present work, we point out that in the conventional
superconductors, the phase with both finite resistivity and gap can also be
realized by tuning the magnetic flux. Nevertheless, to realize this
special phase, the emergence point {\small{$\omega_D/k_F$}} of this phase
must be smaller than the critical point $v_c$ at which the superconducting
gap becomes zero. Thus, small Debye frequency and low
temperature are necessary. Consequently, 
materials Pb, Hg and V, which possess small Debye frequency,\cite{Pb2} are the
some ideal candidates. For the experimental detection, the finite resistivity can be
detected through the electrical methods\cite{HTSC1,HTSC2,HTSC3,SDSC2,SDSC3,SDSC4} whereas the finite gap can be measured by
using the scanning tunneling microscope\cite{HTSC1,HTSC2,HTSC5,SDSC1,SDSC2,SDSC4,STM} or angle-resolved photo-emission
spectroscopy.\cite{HTSC1,ARPES} 

\subsection{Optical response}

We next study the optical
response in the superconducting states in both linear and non-linear regimes.
Properties of the optical current and excited Higgs mode are addressed. 

\subsubsection{Solution of density matrix}

In the optical response, we first choose a specific gauge with zero
superconducting phase for the convenience of the physical analysis, and considering 
the translational symmetry, the spatial gradient terms in Eq.~(\ref{KE}) can be
neglected. Then, the kinetic equation reads
\begin{eqnarray}
&&\partial_T\rho_{\bf k}+i\left[\left(\xi_k+\mu_{\rm
  eff}\right)\tau_3+|\Delta|\tau_1,\rho_{\bf
  k}\right]+i\left[\frac{p^2_s}{2m}\tau_3,\rho_{\bf
  k}\right]\nonumber\\
&&\mbox{}+\frac{1}{2}\{{e{\bf E}}\tau_3,{\bm \partial}_{\bf
  k}\rho_{\bf k}\}+\{{\bf p}_s|\Delta|\tau_2,{\bm \partial}_{\bf
  k}\rho_{\bf k}\}\nonumber\\
&&\mbox{}+\frac{i}{2}\left[{\bf p}_s{\bf p}_s|\Delta|\tau_1,{\bm \partial}_{\bf
  k}{\bm \partial}_{\bf
  k}\rho_{\bf k}\right]=\partial_t\rho_{\bf k}\Big|_{\rm sc}.\label{O1}
\end{eqnarray}
Here, the superconducting momentum ${\bf p}_s=-e{\bf
  A}+\frac{1}{2}{\bm \nabla}_{\bf R}\psi$ and the effective chemical potential ${\mu}_{\rm
  eff}=e\phi+\frac{1}{2}\partial_t\psi$, related by the acceleration relation
$e{\bf E}=\partial_t{\bf p}_s-{\bm \nabla}\mu_{\rm eff}$, are gauge-invariant physical
quantities.\cite{gi0,GOBE1} Particularly, in the presence of the translational
symmetry, the electric field reads $e{\bf E}=\partial_{t}{\bf p}_s-{\bm
  \nabla}{\mu_{\rm eff}}=i\omega{\bf p}_s$ in the optical response with $\omega$
being optical frequency. On the left-hand 
side of Eq.~(\ref{O1}), the third term represents the
Anderson-pseudospin pump effect\cite{Higgs20,Higgs21,Higgs4,Higgs5,Higgs50,Higgs6,Higgs7,Higgs8,Higgs10} and
the forth one is the drive effect, exactly as those revealed in the previous
work by Yu and Wu.\cite{GOBE1} Whereas the last two terms on the left-hand 
side of Eq.~(\ref{O1}), which stand for the Ginzburg-Landau kinetic effect, are
absent in Ref.~\onlinecite{GOBE1}.  

To obtain the solution, we transform Eq.~(\ref{O1}) from the particle space into
the quasiparticle one as: 
\begin{eqnarray}
&&\partial_T\rho^q_{\bf k}+i\left[E_k\tau_3,\rho^q_{\bf
    k}\right]+i\left[\mu_{\rm eff}t_3,\rho^q_{\bf k}\right]+i\left[\frac{p^2_s}{2m}t_3,\rho^q_{\bf k}\right]\nonumber\\
&&\mbox{}+\frac{1}{2}\left\{{e{\bf E}}t_3+2{\bf p}_s|\Delta|t_2,{\bm \partial}_{\bf
  k}\rho^q_{\bf k}+[U^{\dagger}_k{\bm \partial}_{\bf
  k}U_k,\rho^q_{\bf k}]\right\}\nonumber\\
&&\mbox{}+\frac{i}{2}\left[{\bf p}_s{\bf p}_s|\Delta|\tau_1,{\bm \partial}_{\bf
  k}{\bm \partial}_{\bf
  k}\rho_{\bf k}+2[U^{\dagger}_k{\bm \partial}_{\bf
  k}U_k,{\bm \partial}_{\bf
  k}\rho^q_{\bf k}]\right]\nonumber\\
&&\mbox{}+\frac{i}{2}\left[{\bf p}_s{\bf p}_s|\Delta|\tau_1,U^{\dagger}_k{\bm \partial}_{\bf
  k}{\bm \partial}_{\bf
  k}U_k\rho^q_{\bf k}+\rho^q_{\bf k}({\bm \partial}_{\bf
  k}{\bm \partial}_{\bf
  k}U^{\dagger}_k)U_k\right]\nonumber\\
&&\mbox{}-\frac{i}{2}\left[{\bf p}_s{\bf p}_s|\Delta|\tau_1,2U^{\dagger}_k{\bm \partial}_{\bf
  k}U_k\rho^q_{\bf k}U^{\dagger}_k{\bm \partial}_{\bf
  k}U_k\right]=\partial_t\rho^q_{\bf k}\Big|_{\rm sc},~~~~~\label{O2}
\end{eqnarray}
in which $t_i=U^{\dagger}_k\tau_iU_k$.

Then, from Eq.~(\ref{O2}), the solution
of the density matrix in the quasiparticle space is derived
as (refer to Appendix~\ref{ad})
\begin{equation}
\rho^q_{\bf k}=\rho^{q0}_{\bf k}-({\bf k}{\cdot}{\bf v}_s)\rho^{q1}_{\bf
  k}+\frac{({\bf k}{\cdot}{\bf v}_s)^2}{2}\rho^{q2}_{\bf
  k}+mv^2_s\rho^{q3}_{\bf k}+\delta\rho^{qs}_{\bf
  k},\label{or}
\end{equation}
\begin{eqnarray}
&&\rho^{q0}_{\bf k}=\left(\begin{array}{cc}
f(E^+_{\bf k})& 0\\
0 &f(E^-_{\bf k}) 
\end{array}\right),\\
&&\rho^{q1}_{\bf
  k}=\frac{\rho_{m{\bf k}}\tau_0}{4\varepsilon_{k_F}},\\
&&\delta\rho^{qs}_{\bf
  k}=-\frac{({\bf k}{\cdot}{\bf v}_s)}{i\omega\tau_k}\theta\left(\frac{kv_s}{E_k}\right)(\partial_{E_k}\rho^{q0}_{{\bf k}3}+{\hat
  O_k}f_{\bf k})\tau_0,
\end{eqnarray}
in which, {\small{${\bf v}_s=-\frac{e{\bf E}}{i{\omega}m}$}} (refer to Appendix); {\small{${\hat
      O_k}=4u_k^2v_k^2(1/{E_k}-\partial_{E_k})$}} and 
{\small{$f_{\bf k}=[3f(E^+_{\bf k})-3f(E^-_{\bf k})-f(E^+_{\bf
      k}+2E_k)+f(E^-_{\bf k}-2E_k)]/{8}$}}; the specific expressions of
$\rho^{q2}_{\bf k}$ and $\rho^{q3}_{\bf k}$ are given by Eqs.~(\ref{ad7}) and
(\ref{ad5}) in Appendix, respectively. 

As seen from Eq.~(\ref{or}), the first term in $\rho^q_{\bf k}$ represents the
quasiparticle distribution. The second term, in which
$\rho_{m{\bf k}}$ is exactly same as the Meissner-superfluid density
[Eq.~(\ref{rmk})] in the stationary magnetic response, stands for the Meissner
response. The third and forth terms denote the
nonlinear response. The last term is the
scattering contribution, which emerges at $kv_s>E_k$ as mentioned in Sec.~\ref{tfm}. 

\subsubsection{Optical current}

We first investigate the properties of the optical current. In contrast to the
two-fluid model in the literature,\cite{Ba0,Ba1,Ba4,L6,L7,L9,NL1,NL2,two} we
show that the optical current is well captured by the three-fluid model
described in Sec.~\ref{tfm}. Specifically, with Eqs.~(\ref{or}) and
(\ref{current}), by neglecting the 
nonlinear response,  the optical current reads:  
\begin{eqnarray}
{\bf j}&=&\frac{2e}{m}\sum_{\bf k}{\bf k}\rho_{{\bf k}0}=\frac{2e}{m}\sum_{\bf k}{\bf k}\rho^q_{{\bf k}0}\nonumber\\
&=&\frac{2e}{m}\sum_{\bf k}{\bf k}[\rho^{q0}_{{\bf
  k}0}-({\bf k}{\cdot}{\bf v}_s)\rho^{q1}_{{\bf
  k}}+\delta\rho^{qs}_{\bf k}].\label{oc1}
\end{eqnarray}

At $v_s<{v_L}$ with only the non-viscous superfluid, the current is
written as
\begin{equation}
{\bf j}=\frac{e^2{\bf E}}{im\omega}\sum_{\bf k}\cos^2\theta_{\bf k}[\rho_{m{\bf
    k}}-4\varepsilon_{k_F}\partial_{E_k}f(E_k)]. 
\end{equation}
Besides the Meissner supercurrent  ($\rho_{m{\bf  
    k}}$), there also exists Bogoliubov quasiparticle current
[$4\varepsilon_{k_F}\partial_{E_k}f(E_k)$] in the superfluid during the optical response.  
The presence of the Bogoliubov quasiparticle current is natural, since the drive from
the optical field causes the drift of the electron states, resulting a
center-of-mass momentum in superconducting states.\cite{GOBE1} 

As for the case $v_s>{v_L}$ with the presences of the normal fluid, non-viscous and viscous
superfluids, by using the same expansion of $f(E^{\pm}_{\bf k})$ in each regions in Sec.~\ref{ECIM}, in the weak scattering limit,
  the current becomes 
\begin{equation}
{\bf j}=({\sigma}_{\rm P_{nv}}+{\sigma}_{\rm P_{v}}+{\sigma}_{\rm U}){\bf E},\label{oc3}
\end{equation}
with 
\begin{eqnarray}
{\sigma}_{\rm P_{nv}}&=&\frac{e^2}{im\omega}\sum_{{\bf k}\in{\rm P_{nv}}}\cos^2\theta_{\bf k}[\rho_{m{\bf
    k}}-4\varepsilon_{k_F}\partial_{E_k}f(E_k)],\label{ocf1}\\
{\sigma}_{\rm P_{v}}&=&\frac{e^2}{m}\sum_{{\bf k}\in{\rm P_{v}}}\cos^2\theta_{\bf k}\Bigg[\frac{\rho_{m{\bf
    k}}}{i\omega+(2\tau_k)^{-1}}-\frac{4\varepsilon_{k_F}\partial_{E_k}f(E_k)}{i\omega+\tau_k^{-1}}\Bigg],\nonumber\\
&&~~\label{ocf2}\\
{\sigma}_{\rm U}&=&-\frac{e^2}{m}\sum_{{\bf k}\in{\rm
    U}}\cos^2\theta_{\bf
  k}\frac{4\varepsilon_{k_F}\partial_{\xi_k}f(\xi_k)}{i\omega+\tau_k^{-1}}. \label{ocf3}
\end{eqnarray}
Specifically, the excited superfluid current consists of the Meissner
supercurrent ($\rho_{m{\bf k}}$) and Bogoliubov 
quasiparticle current [$4\varepsilon_{k_F}\partial_{E_k}f(E_k)$], as mentioned
above. Due to the presence of the friction between the superfluid and
normal-fluid currents mentioned in Sec.~\ref{tfm}, the superfluid current is
separated into the non-viscous ${\sigma}_{\rm P_{nv}}{\bf E}$ and viscous
${\sigma}_{\rm P_{v}}{\bf E}$ ones, and the former (latter) exhibits zero
(finite) resistance $\tau_k^{-1}$. Whereas the normal-fluid optical conductivity
${\sigma}_{\rm U}$ exhibits the well-known Drude-model behavior. Particularly,
in the normal state with the normal fluid alone, it exactly reduces to the one 
$\sigma_{\rm U}=\frac{e^2N\tau}{m(1+i\omega\tau)}$ from Drude model.\cite{Drude} 

In the superconducting state, at low temperature, few  Bogoliubov quasiparticles
[$f(E_k)\approx0$] are excited in the superfluid. Thus, the Bogoliubov
quasiparticle current is marginal and the superfluid current only consists of 
Meissner supercurrent. In this situation, if we neglect the friction
between superfluid and normal-fluid currents, i.e., the viscous superfluid
(${\sigma}_{\rm P_{v}}$), the optical conductivity $\sigma=\sigma_{\rm
  U}+\sigma_{\rm P_{nv}}$ from our theory
[Eqs.~(\ref{ocf1}) and (\ref{ocf3}) with $f(E_k)\approx0$] is exactly same as
the one $\sigma_{\rm
  two}=\frac{e^2\rho_m}{im\omega}+\frac{e^2\tau\rho_n}{m(1+i\omega\tau)}$ from
two-fluid model,\cite{Ba0,Ba1,Ba4,L6,L7,L9,NL1,NL2,two} in which $\rho_m$ is
the total Meissner-superfluid density and $\rho_n$ denotes the total
normal-fluid density. Nevertheless, the presence of viscous superfluid here suggests
that the optical response is captured by the three-fluid model and the two-fluid
model in the literature\cite{Ba0,Ba1,Ba4,L6,L7,L9,NL1,NL2,two} is insufficient
for a complete picture. Actually, although the viscous superfluid has been
hinted in the stationary magnetic response in the
literature,\cite{MBm,Ba1} it has long been overlooked in the optical
response.

\subsubsection{Higgs mode}
\label{Higgs}
Finally, we discuss the optically excited Higgs mode. Comparison between the
Anderson-pseudospin pump effect\cite{Higgs20,Higgs21,Higgs4,Higgs5,Higgs50,Higgs6,Higgs7,Higgs8,Higgs10} [third
term in  Eq.~(\ref{O1})] and the drive effect [forth term in
Eq.~(\ref{O1})] revealed in the previous theory\cite{GOBE1} by Yu and Wu is
addressed. Particularly, in Ref.~\onlinecite{GOBE1}, it is reported that in the
excitation of the Higgs mode, the drive effect is the dominant effect and the
pump effect is marginal. Nevertheless, as pointed out in Sec.~\ref{DFKE}, the
Ginzburg-Landau kinetic-energy terms [seventh and eighth terms in
Eq.~(\ref{O1})] are absent in Ref.~\onlinecite{GOBE1}. With these two terms, we
show that the previous conclusion in Ref.~\onlinecite{GOBE1} only holds at
finite temperature. 

Specifically, with the
solution of density matrix in the optical response [Eq.~(\ref{or})], the gap 
equation [Eq.~(\ref{ge})] becomes
\begin{eqnarray}
&&\Delta=V\sum_{{\bf k}\in{\rm P}}\frac{\Delta}{E_k}\Bigg\{-a_k+\frac{({\bf k}\cdot{\bf
    v}_s)^2\xi^2_k}{2(\omega^2-E^2_k)}\Bigg[\Bigg(1-\frac{\omega|\Delta|^2}{E^3_k}\Bigg)\partial_{E_k}\nonumber\\  
&&\mbox{}+\frac{1}{2\varepsilon_{k_F}}\left(1+\frac{\omega|\Delta|^2}{\xi^2_kE_k\cos^2\theta_{\bf
    k}}\right)\Bigg]\partial_{E_k}f(E_k)+\varepsilon_{p_s}a_k\frac{|\Delta|^2}{E^2_k}\nonumber\\
&&\mbox{}\times\frac{E_k\cos^2{\theta_{\bf 
    k}}-\omega}{\omega^2-E^2_k}+\frac{\varepsilon_{p_s}\omega{a_k}}{\omega^2-E_k^2}\Bigg\},\label{H1}
\end{eqnarray}
in which $a_k=f(E_k)-1/2$. The above gap equation is calculated in the pairing
region alone. In principle, the superconducting gap is self-consistently
determined by the above gap equation. Nevertheless, for a weak optical field at
low temperature, one has $|\delta\Delta|=|\Delta_0-\Delta|{\ll}|\Delta_0|$ with
$\Delta_0=V\sum_{{\bf k}\in{\rm
    P}}\frac{1}{2E_k}$ being the gap at zero temperature. 

At low temperature, considering the large Fermi energy in conventional superconductors,
from Eq.~(\ref{H1}), $\delta\Delta$ reads
\begin{equation}
\delta\Delta=\Delta_0V\sum_{{\bf k}\in{\rm
    P}}\frac{f(E_k)}{E_k}-\delta\Delta^{\rm pump}-\delta\Delta^{\rm drive}, \label{g3}
\end{equation}
with
\begin{eqnarray}
\frac{\delta\Delta^{\rm pump}}{V}&=&\varepsilon_{p_s}\sum_{{\bf k}\in{\rm
    P}}\frac{\omega{a_k}\Delta_0}{E_k(\omega^2-E_k^2)} \label{pump} \\
\frac{\delta\Delta^{\rm drive}}{V}&=&\varepsilon_{p_s}\Delta_0\sum_{{\bf k}\in{\rm
    P}}\Bigg[2\varepsilon_{k_F}\frac{\xi^2_k\cos^2\theta_{\bf k}\partial^2_{E_k}f(E_k)}{E_k(\omega^2-E^2_k)}\nonumber\\
&&\mbox{}\times\left(1-\frac{\omega|\Delta_0|^2}{E^3_k}\right)+a_k\frac{|\Delta|^2}{E^3_k}\frac{E_k\cos^2{\theta_{\bf 
    k}}-\omega}{\omega^2-E^2_k}\Bigg]. \nonumber\\
\label{drive}
\end{eqnarray}
On the right-hand side of Eq.~(\ref{g3}), the first term directly leads to the
decrease of the superconducting gap as a consequence of the thermal
effect. Particularly, this term is finite after the THz pulse and hence causes a
plateau of the superconducting gap, in consistency with the experimental
findings.\cite{NL7,NL8} 
We point out that the second term comes from the Anderson-pseudospin pump
effect.\cite{Higgs20,Higgs21,Higgs4,Higgs5,Higgs50,Higgs6,Higgs7,Higgs8,Higgs10} The third term arises from the 
drive effect.\cite{GOBE1} Both effects in the excitation of the Higgs mode,
proportional to $\varepsilon_{p_s}=p^2_s/(2m)$, oscillate at twice optical
frequency. 

By comparing the relative contribution of these two effects, near the Fermi surface, 
at zero temperature in the absence of thermal effect, 
the ratio between the drive and pump effects is $r_{{\rm
      drive}/{\rm pump}}\approx\Big|\frac{\Delta-3\omega}{3\omega}\Big|$, and  
in the THz regime, both effects contribute. Whereas at finite temperature, 
thanks to the large Fermi energy, the drive effect [Eq.~(\ref{drive})] becomes
\begin{eqnarray}
\frac{\delta\Delta^{\rm drive}}{V}&=&\varepsilon_{p_s}\Delta_02\varepsilon_{k_F}\sum_{{\bf k}\in{\rm
    P}}\Bigg[\frac{\xi^2_k\cos^2\theta_{\bf
          k}\partial^2_{E_k}f(E_k)}{E_k(\omega^2-E^2_k)}\nonumber\\
&&\mbox{}\times\left(1-\frac{\omega|\Delta_0|^2}{E^3_k}\right)\Bigg].
\end{eqnarray}
Then, one finds $r_{{\rm
    drive}/{\rm
    pump}}\approx\Big|\frac{\varepsilon_{k_F}}{6\omega}\frac{(\omega-|\Delta|)|\Delta|}{T^2_{\rm
    eff}\cosh^3(\frac{|\Delta|}{2T_{\rm eff}})}\Big|$, and hence, the drive effect
plays a dominant role in the excitation of the Higgs mode.  

Actually, the dominant role of the drive effect can also be understood as
follows. It is noted that at low frequency and small order parameter, 
 the drive effect [Eq.~(\ref{drive})] becomes
\begin{equation}
{\delta\Delta^{\rm drive}}/{V}=-2\varepsilon_{p_s}\Delta_0\lambda=-mv^2_s\Delta_0\lambda,
\end{equation}
which is exactly the kinetic-energy term in the Ginzburg-Landau equation [first term in
Eq.~(\ref{MGL})]. Consequently, the drive effect in our microscopic theory is
related to the kinetic energy in the Ginzburg-Landau theory, in which the vector
potential is involved as $({\bf k}_F\cdot{\bf A})^2/m^2=4\varepsilon_{k_F}{\bf
  A}^2/(2m)$ at finite temperature. Nevertheless, in the pump effect, the vector
potential is involved as ${\bf A}^2/(2m)$. These two responses of the
vector potentials are totally different, and thanks to the large Fermi energy,
the drive effect makes the dominant contribution. Consequently, the
Liouville\cite{Higgs20,Higgs21,Higgs5} or
Bloch\cite{Higgs4,Higgs50,Higgs6,Higgs7,Higgs8,Higgs10} equation in the
literature with the pump effect alone is insufficient to study the optical excitation of
the Higgs mode.  However, although the deficiency of the
Liouville or Bloch equation has been hinted according to the Ginzburg-Landau
theory, it has long been overlooked in the study of the Higgs mode in the literature.

Particularly, in the experiments for the detection of Higgs
mode,\cite{NL5,NL7,NL8,NL9,NL10} the thermal effect is inevitable because of the
intense THz field. This conclusion is supported by the experimentally discovered
plateau of the superconducting gag after the THz pulse, which is attributed to
the thermal effect as mentioned above. Therefore, we believe that the
experimentally observed excitation of the Higgs mode is dominated by the drive
effect. This conclusion is also supported by our numerical calculation (refer to Appendix~\ref{ae}).

\section{SUMMARY AND DISCUSSION}
\label{summary}

In summary, we extend the kinetic theory by Yu and Wu\cite{GOBE1} to include the 
superfluid, so that both the normal-fluid and superfluid dynamics are
involved. As a gauge-invariant theory for the 
electromagnetic response, our kinetic equation can be applied to both the
magnetic and optical responses. We first focus on the weak-scattering case
in the present work. Rich physics is revealed. 

Specifically, in the electromagnetic response, we show that the superconducting 
velocity $v_s$ is always excited by the electromagnetic field. Particularly, 
a threshold $v_L=|\Delta|/k_F$ of superconducting velocity $v_s$ for the
emergence of normal fluid and hence the scattering is predicted from our theory,
i.e., the normal fluid and scattering appear only when $v_s>v_L$. We refer to
this threshold as Landau threshold, following Landau in bosonic liquid helium II
theory.\cite{Landau} 
Interestingly, we find that there also exists friction between the normal-fluid
and superfluid currents. Due to this friction,
part of superfluid becomes viscous. Therefore, the superfluid consists of non-viscous
superfluid and viscous one. Consequently, we propose a
three-fluid model at $v_s{\ge}v_L$: normal fluid, non-viscous and viscous
superfluids. We show that from this three-fluid model, the physics of the
electromagnetic response in the superconducting states can be well captured. 

The physical picture behind these predictions can be understood as follows. 
At a small superconducting velocity, the superconducting state behaves like the 
BCS state, in which all particles in the momentum spherical shell participate in
the pairing. Thus, there only exists superfluid.  
In the case of a large superconducting velocity with $v_s>v_L$, as revealed in
the previous works,\cite{FF7,FF8,FF9,FF1} besides the
pairing region, there also exists unpairing (U) region, in which the
particles no longer participate in the pairing and behave like the normal particles. 
Hence, both the normal fluid (from U region) and superfluid (from
pairing region) are present. Furthermore, we find that there exists a special
region (P$_{\rm v}$ region) in the pairing region which share the  
same momentum magnitude with U region. Particles in this P$_{\rm v}$ region
participate in the pairing but experience the scattering with those in U region due
to the isotropic short-range impurity scattering in conventional superconducting
metals, leading to the friction between the superfluid and normal-fluid
currents. Consequently, the superfluid in P$_{\rm v}$ region becomes
viscous. Whereas the superfluid in the remaining pairing region (P$_{\rm nv}$
region) is still non-viscous.  

For the stationary magnetic response, in the case with $v_s<v_L$ in which
only the non-viscous superfluid is present, we prove that the excited superfluid current is
the Meissner supercurrent and near the critical temperature, the gap equation in our
theory reduces to the Ginzburg-Landau equation.\cite{Ginzburg} As for the
situation with $v_s{\ge}v_L$ where both the superfluid and
normal fluid are present, differing from the excited Meissner supercurrent in
the superfluid, no current is directly excited from the magnetic flux in the normal
fluid. Nevertheless, through the friction drag with superfluid
current, the normal-fluid current is induced. Moreover, thanks to this friction,
the superfluid is separated into the non-viscous (from P$_{\rm nv}$
region) and viscous (from  P$_{\rm v}$ region) ones. Thus, the stationary
magnetic response is captured by the three-fluid model. Moreover, because of the
normal-fluid and viscous-superfluid currents, the penetration depth is
influenced by the scattering. 
Particularly, by only considering the viscous superfluid current, the dependence of
penetration depth on mean free path from our theory is exactly same
as the one from Tinkham's discussion.\cite{Ba1} Nevertheless, since there
also exists the normal-fluid current induced by friction drag and non-viscous
superfluid current, an extension of penetration depth is proposed. 

In addition, when $v_s{\ge}v_L$, a modified Ginzburg-Landau equation
is revealed, in which the calculation of the phenomenological parameters are
restricted to the pairing region. Furthermore, at $v_s>\omega_D/k_F$
before the superconducting gap is destroyed, we predict an exotic phase, in
which the non-viscous superfluid vanishes, leaving only the viscous superfluid
and normal fluid. Thus, interestingly, this phase shows the finite resistivity
but with a finite superconducting gap. Actually, in high-temperature
superconductors\cite{HTSC1,HTSC2,HTSC3,HTSC4,HTSC5,HTSC6} and strongly
disordered superconductors,\cite{SDSC1,SDSC2,SDSC3,SDSC4} the phase with both
finite resistivity and gap, known as pseudogap phase, has been widely studied. 
We point out that in the conventional superconductors, the phase with both
finite resistivity and gap can also be realized by tuning the magnetic flux.

As for the optical response, the excited superconducting ${v_s}$ 
oscillates with time. When $v_s<v_L$, only the non-viscous superfluid is present whereas
at $v_s{\ge}v_L$, there exist normal fluid (from U region), non-viscous (from
P$_{\rm nv}$ region) and viscous (from P$_{\rm v}$ region) superfluids. 
We show that in the optical
response, the excited normal-fluid current exhibits the Drude-model behavior as
it should be. Whereas in the superfluid, we find that the superfluid current is
excited and it consists of the Meissner supercurrent, which has the same form as
that in the magnetic response, as well as the Bogoliubov quasiparticle current.
Particularly, at low temperature, few Bogoliubov quasiparticles are
excited in the pairing region and hence the Bogoliubov quasiparticle current is
marginal. Then, the normal-fluid current and the superfluid current
which only consists of Meissner supercurrent 
are exactly same as those in the original two-fluid
model.\cite{Ba0,Ba1,Ba4,L6,L7,L9,NL1,NL2,two} However, there exists
friction between the superfluid and normal-fluid currents, and due to this
friction, the superfluid is separated into the non-viscous and viscous
ones. The presence of viscous superfluid suggests that the
optical response is also captured by the three-fluid model and the two-fluid
model\cite{Ba0,Ba1,Ba4,L6,L7,L9,NL1,NL2,two} in the literature is insufficient
for a complete picture. Actually, although the viscous superfluid has
been hinted in the stationary magnetic response in the literature,\cite{MBm,Ba1}
it has long been overlooked in the optical response. 

Based on the three-fluid model, the expression of the optical conductivity is
revealed. We also give the expression of the optical excitation of the Higgs mode. 
By comparing the contributions from the drive and Anderson-pseudospin pump
effects, we find that the drive effect is dominant at finite temperature whereas
at zero temperature, both effects contribute. Actually, the drive
effect in our microscopic theory is related to the kinetic energy in the
Ginzburg-Landau theory, in which
the vector potential is involved as $({\bf k}_F\cdot{\bf
  A})^2/m^2=4\varepsilon_{k_F}{\bf A}^2/(2m)$ at finite
temperature. Nevertheless,  in the pump effect, the vector
potential is involved as ${\bf A}^2/(2m)$. These two responses of the
vector potentials are totally different, and thanks to the large Fermi energy,
the drive effect makes the dominant contribution. Consequently, the
Liouville\cite{Higgs20,Higgs21,Higgs5} or Bloch\cite{Higgs4,Higgs50,Higgs6,Higgs7,Higgs8,Higgs10} equation in the literature 
with the pump effect alone is insufficient to study the optical excitation of
the Higgs mode.  However, although the deficiency of the
Liouville or Bloch equation has been hinted according to the Ginzburg-Landau
theory, it has long been overlooked in the study of the Higgs mode in the
literature. Particularly, in the experiments for
the detection of Higgs mode,\cite{NL7,NL8,NL9} since the thermal effect is
inevitable because of the intense THz field, we believe that the experimentally 
observed excitation of the Higgs mode is dominated by the drive effect.

Finally, we discuss the charge density in the superconducting state from the
dynamic viewpoint. In the superfluid, from the BCS theory, the charge
density with momentum ${\bf k}$ reads\cite{GOBE1,cn0,cn1,cn2,cn3} 
\begin{equation}
en_{\bf
  k}=e\sum_{\sigma}\langle{c^{\dagger}_{{\bf 
      k}\sigma}c_{{\bf k}\sigma}}\rangle=2ev^2_k+2e\frac{\xi_k}{E_k}f(E_k),
\end{equation} 
consisting of the charge densities of the condensate\cite{cn0,cn1,cn2,cn3}
{\small{$2ev^2_k$}} and Bogoliubov quasiparticles\cite{cn0,cn1,cn2,cn3,cn4,cn5}
{\small{$2e\frac{\xi_k}{E_k}f(E_k)$}}. Whereas in the normal state, one has $en_{\bf
  k}=2ef(\xi_k)$. Therefore, there exists the charge-density difference
between the superconducting and normal states, which is related to the well-known
particle-number unconservation in the BCS theory. 
Interestingly, we find that this charge-density difference can be compensated by
the Meissner-superfluid density $\rho_{m{\bf k}}$ [Eq.~(\ref{rmk})] as 
\begin{equation}
\label{ndk}
2ef(\xi_k)=2ev^2_k+2e\frac{\xi_k}{E_k}f(E_k)-eC_k\rho_{m{\bf k}}+eO(|\Delta|^4),
\end{equation}
with a prefactor $C_k=D_0\xi_k/(3N_0)$, guaranteeing the charge-density
conservation in the superconducting states. As seen from the right-hand
side of above equation, in addition to the condensate and Bogoliubov
quasiparticles, the charge density in the superconducting states also consists
of the contribution from the Meissner density $\rho_{m{\bf k}}$. At zero
temperature, as the Bogoliubov quasiparticles, i.e., thermal excitations,
vanish, what remain are the condensate from the BCS ground state and the
Meissner charge fluctuation on top of the condensate. By noticing that all the
electromagnetic responses in superconductors at zero temperature come from the
Meissner current, one can draw the conclusion that only the Meissner charge
fluctuation contributes to the superconducting response and the condensate simply
provides a rigid background. This is in contrast to the previous
textbook understanding\cite{Ba3,U2,FF6,cm1,cm2,cm3,cm4} that the 
supercurrent is a collective motion of the condensate.\cite{note} 

\begin{acknowledgments}
This work was supported by the National Natural Science Foundation of 
China under Grants No.\ 11334014 and No.\ 61411136001.  
\end{acknowledgments}

\begin{appendix}

\section{Derivation of Eq.~(\ref{s5})}
\label{aa}

In this section, we derive Eq.~(\ref{s5}). Specifically, by taking the impurity
scattering as the short-range one, i.e., $|V_{\bf kk'}|^2\approx|V_0|^2$, after
the integration over the angle in Eq.~(\ref{s2}), one obtains 
\begin{widetext}
\begin{eqnarray}
&&\partial_t\rho^q_{\bf k}\Big|_{\rm sc}=-n_i\pi{D_0}|V_0|^2\int{\frac{d\xi_{k'}}{k_Fv_s}}(1-\eta_{kk'})\left(\begin{array}{cc}
(\rho^{q}_{{\bf k},11}-\rho^q_{{\bf k}',11})\Big|_{\cos\theta_{\bf
    k'}=\frac{E^+_{\bf k}-E_{k'}}{k_Fv_s}} & 0\\
0& (\rho^{q}_{{\bf k},22}-\rho^q_{{\bf k}',22})\Big|_{\cos\theta_{\bf
    k'}=\frac{E^-_{\bf k}+E_{k'}}{k_Fv_s}}
\end{array}\right)\nonumber\\
&&\mbox{}-n_i\pi{D_0}|V_0|^2\int{\frac{d\xi_{k'}}{k_Fv_s}}(1+\eta_{kk'})\left(\begin{array}{cc}
(\rho^{q}_{{\bf k},11}-\rho^q_{{\bf k}',22})\Big|_{\cos\theta_{\bf
    k'}=\frac{E^+_{\bf k}+E_{k'}}{k_Fv_s}} & 0\\
0& (\rho^{q}_{{\bf k},22}-\rho^q_{{\bf k}',11})\Big|_{\cos\theta_{\bf
    k'}=\frac{E^-_{\bf k}-E_{k'}}{k_Fv_s}}
\end{array}\right).\label{aa1}
\end{eqnarray}\\
\end{widetext}
Thanks to the large Fermi energy, we approximately take the emergence of the
scattering around the Fermi surface by setting $|\xi_{k}|,|\xi_{k'}|<E_c$ in
Eq.~(\ref{aa1}). Then, one has $|\xi_k|-E_c<\xi_{k'}<|\xi_k|+E_c$. By using
the mean value theorem for integrals, Eq.~(\ref{aa1}) becomes 
\begin{widetext}
\begin{eqnarray}
&&\partial_t\rho^q_{\bf k}\Big|_{\rm sc}=-2n_i\pi{D_0}|V_0|^2\lambda_c\frac{\xi_k(\xi_k+\xi_{k'})}{E^2_k}\left(\begin{array}{cc}
(\rho^{q}_{{\bf k},11}-\rho^q_{{\bf k}',11})\Big|^{|\xi_k|=|\xi_{k'}|}_{\cos\theta_{\bf
    k'}=\cos\theta_{\bf k}} & 0\\
0& (\rho^{q}_{{\bf k},22}-\rho^q_{{\bf k}',22})\Big|^{|\xi_k|=|\xi_{k'}|}_{\cos\theta_{\bf
    k'}=\cos\theta_{\bf k}}
\end{array}\right)\nonumber\\
&&\mbox{}-2n_i\pi{D_0}|V_0|^2\lambda_c\left(1+\frac{|\Delta|^2}{E^2_k}\right)\left(\begin{array}{cc}
(\rho^{q}_{{\bf k},11}-\rho^q_{{\bf k}',22})\Big|^{|\xi_k|=|\xi_{k'}|}_{\cos\theta_{\bf
    k'}-\cos\theta_{\bf k}=\frac{2E_{k}}{k_Fv_s}} & 0\\
0& (\rho^{q}_{{\bf k},22}-\rho^q_{{\bf k}',11})\Big|^{|\xi_k|=|\xi_{k'}|}_{\cos\theta_{\bf
    k'}-\cos\theta_{\bf k}=-\frac{2E_{k}}{k_Fv_s}}
\end{array}\right),\label{aa2}
\end{eqnarray}\\
\end{widetext}
with the dimensionless parameter $\lambda_c=2E_c/(k_Fv_s)$. It is noted that the 
first term on the right-hand side of Eq.~(\ref{aa2}) is zero as a consequence of
the particle-hole symmetry under the particle-hole transformation\cite{KB3}
$\xi_{k}\rightarrow-\xi_{k}$. Then, Eq.~(\ref{s5}) is obtained.

\section{Derivation of Eq.~(\ref{mr})}
\label{ab}

We derive Eq.~(\ref{mr}) in this part. Considering the large Fermi
energy in conventional superconductors, one can neglect $\varepsilon_{{\bf p}-2e{\bf 
    A}}$ on the left-hand side of Eq.~(\ref{E7}). Then, by using Eq.~(\ref{E7})
to substitute $\rho_{{\bf k}\pm}$ in Eq.~(\ref{E8}), one has
\begin{eqnarray}
&&\frac{\bf k}{m}\cdot{\bm \nabla}\rho_{{\bf k}0}=\frac{{\bm \nabla}|\Delta|^2\cdot{\bm \partial}_{{\bf
      k}}\rho_{{\bf k}0}}{2\xi_k}+\frac{({\bm \nabla}|\Delta|^2\cdot{\bm \partial}_{{\bf
      k}})({\bf p}_s\cdot{\bm \partial}_{\bf k})\rho_{{\bf k}3}}{2\xi_k}\nonumber\\
&&\mbox{}-\frac{({\bm \nabla}|\Delta|^2\cdot{\bm \partial}_{{\bf
      k}})({\bf p}_s\cdot{\bm \partial}_{\bf k})(\rho_{{\bf
      k}3}/\xi_k)}{2}+\left\{\partial_{t}\rho_{\bf k}\Big|_{\rm sc}\right\}_3.\label{ab1}
\end{eqnarray}
In the quasiparticle space, Eq.~(\ref{ab1}) becomes
\begin{eqnarray}
&&\frac{\bf k}{m}\cdot{\bm \nabla}\rho^q_{{\bf k}0}={\bm \nabla}|\Delta|^2\cdot\Bigg[\frac{{\bm \partial}_{{\bf
      k}}\rho^q_{{\bf k}0}}{2\xi_k}+\frac{{\bm \partial}_{{\bf
      k}}({\bf p}_s\cdot{\bm \partial}_{\bf k})}{2\xi_k}\left(\frac{\xi_k\rho^q_{{\bf k}3}}{E_k}\right)\nonumber\\
&&\mbox{}-\frac{{\bm \partial}_{{\bf
      k}}({\bf p}_s\cdot{\bm \partial}_{\bf k})(\rho^q_{{\bf
      k}3}/E_k)}{2}\Bigg]+\frac{\xi_k}{E_k}\left\{\partial_{t}\rho^q_{\bf k}\Big|_{\rm sc}\right\}_3.\label{ab2}
\end{eqnarray}

In the presence of a superconducting momentum ${\bf p}_s=-e{\bf 
  A}+\frac{1}{2}{\bm \nabla}_{\bf R}\psi$, i.e., the center-of-mass momentum, the
superconducting state behaves like the FFLO-like
state.\cite{FF4,FF5,FF6,FF7,FF8,FF9,FF1} 
Consequently, at the weak scattering limit, 
the solution of density matrix reads
\begin{equation}
\rho^q_{\bf k}=\rho^{q0}_{\bf k}+\delta\rho^q_{\bf k}+\delta\rho^{qs}_{\bf k},\label{ab3}
\end{equation}
with
\begin{equation}
\rho^{q0}_{\bf k}=\left(\begin{array}{cc}
f(E^+_{\bf k})& 0\\
0 &f(E^-_{\bf k}) 
\end{array}\right).
\end{equation} 
Here, $\rho^{q0}_{\bf k}$ is the quasiparticle distribution of the FFLO-like state with $E^{\pm}_{\bf
  k}={\bf k}\cdot{\bf v}_s{\pm}E_k$ and ${\bf v}_s={\bf
  p}_s/m$; $\delta\rho^q_{\bf k}$ denotes the disturbance from the FFLO-like state in
the magnetic response in the absence of the scattering; $\delta\rho^{qs}_{\bf
  k}$ represents the scattering contribution. By substituting
Eq.~(\ref{ab3}) into Eq.~(\ref{ab2}), one can construct $\delta\rho^q_{\bf k}$ as  
\begin{eqnarray}
\delta\rho^q_{\bf k}&=&({\bf k}{\cdot}{\bf v}_s)\left(\begin{array}{cc}
a^+_{\bf k}& 0 \\
0 &a^-_{\bf k} 
\end{array}\right)+({\bf k}{\cdot}{\bf v}_s)|\Delta|^2\left(\begin{array}{cc}
m^+_{\bf k}& 0\\
0 &m^-_{\bf k}
\end{array}\right)\nonumber\\
&&+\frac{({\bf k}{\cdot}{\bf v}_s)^2}{2}\left(\begin{array}{cc}
b^+_{\bf k}& 0\\
0 &b^-_{\bf k} 
\end{array}\right),
\end{eqnarray} 
Then, Eq.~(\ref{ab2}) becomes
\begin{eqnarray}
&&\frac{\bf k\cdot{\bm
    \nabla}|\Delta|^2}{m}\left[({\bf k}\cdot{\bf
    v}_s)\frac{m^+_{\bf k}+m^-_{\bf
      k}}{2}+\frac{\partial_{E_k}\rho^q_{k0}}{2E_k}\right]=\frac{{\bf v}_s\cdot{\bm
    \nabla}|\Delta|^2}{2\xi_k} \nonumber\\
&&\mbox{}\times\Bigg[\partial_{E_k}\rho^{q0}_{k3}+\frac{a^+_{\bf k}+a^-_{\bf k}}{2}+({\bf k}\cdot{\bf
    v}_s)\partial_{E_k}\left(\frac{a^+_{\bf k}-a^-_{\bf k}}{2}\right)\nonumber\\
&&\mbox{}+({\bf k}\cdot{\bf
    v}_s)\frac{b^+_{\bf k}+b^-_{\bf k}}{2}\Bigg]+\frac{\bf k}{m}\cdot{\bm
    \nabla}|\Delta|^2\Bigg[\frac{\partial_{E_k}\rho^{q0}_{k0}}{2E_k}+\frac{({\bf k}\cdot{\bf v}_s)}{E_k}\nonumber\\
&&\mbox{}\times\partial_{E_k}\left(\frac{\rho^{q0}_{k3}}{E_k}\right)+\frac{\varepsilon_{p_s}}{\xi_k}\frac{\partial_{E_k}\rho^{q0}_{k0}}{E_k}\Bigg]+\frac{{\bf
  v}_s\cdot{\bm
    \nabla}|\Delta|^2}{2\xi_k}\frac{({\bf k}\cdot{\bf
    v}_s)}{E_k}\nonumber\\
&&\mbox{}\times\partial_{E_k}\rho^{q0}_{k0}+\frac{\rho^q_{{\bf k}3}}{4\xi_kE_k\varepsilon_k}({\bf k}\cdot{\bf
    v}_s)\frac{\bf k}{m}\cdot{\bm
    \nabla}|\Delta|^2,\label{ab4}
\end{eqnarray}
in which we have neglected the terms higher than the second order of $|\Delta|$ or 
$({\bf k}\cdot{\bf v}_s)$. 

Considering the large Fermi energy, one can neglect
$\varepsilon_{p_s}$ term in Eq.~(\ref{ab4}) and obtains
\begin{eqnarray}
&&\frac{\bf k\cdot{\bm
    \nabla}|\Delta|^2}{m}({\bf k}\cdot{\bf
    v}_s)\Bigg[\frac{m^+_{\bf k}+m^-_{\bf
      k}}{2}-\frac{\partial_{E_k}}{E_k}\left(\frac{\rho^{q0}_{k3}}{E_k}\right)-\frac{\rho^q_{{\bf k}3}}{\xi_kE_k}\nonumber\\
&&\mbox{}\times\frac{1}{4\varepsilon_k}\Bigg]-\frac{{\bf v}_s\cdot{\bm
    \nabla}|\Delta|^2}{2\xi_k}\Bigg\{\partial_{E_k}\rho^{q0}_{k3}+\frac{a^+_{\bf k}+a^-_{\bf k}}{2}+({\bf k}\cdot{\bf
    v}_s)\nonumber\\
&&\mbox{}\times\Bigg[\partial_{E_k}\left(\frac{a^+_{\bf k}-a^-_{\bf k}}{2}\right)+\frac{b^+_{\bf k}+b^-_{\bf k}}{2}+\frac{\partial_{E_k}\rho^{q0}_{k0}}{E_k}\Bigg]\Bigg\}=0.~~~~~~\label{ab5}
\end{eqnarray}
It is noted that Eq.~(\ref{ab5}) holds in the entire momentum
space. Consequently, one has
\begin{eqnarray}
&&a^{\pm}_{\bf k}=\mp{\partial}_{E_k}f(E^{\pm}_{\bf k}),\\
&&b^{\pm}_{\bf
  k}={\partial}^2_{E_k}f(E^{\pm}_{\bf k})+\frac{{\partial}_{E_k}f(E^{\pm}_{\bf
  k})}{E_k},\\
&&m^{\pm}_{\bf
    k}=\pm\left[\frac{1}{E_k}\partial_{E_k}+\frac{1}{4\xi_k\varepsilon_k}\right]\frac{f(E^{\pm}_{\bf
    k})}{E_k}.
\end{eqnarray}

As for the scattering contribution $\delta\rho^{qs}_{\bf k}$, one has
\begin{equation}
\frac{\bf k\cdot{\bm \nabla}}{m}\delta\rho^{qs}_{{\bf
    k}0}=\frac{\xi_k}{E_k}\left\{\partial_{t}\rho^q_{\bf k}\Big|_{\rm sc}\right\}_3.\label{ab7} 
\end{equation}
In the weak scattering limit, the scattering only causes the momentum (current)
relaxation. Therefore, by keeping the linear-order terms of $({\bf k}\cdot{\bf
  v}_s)$ in $\rho^q_{\bf k}$, from Eq.~(\ref{s5}), Eq.~(\ref{ab7}) becomes
\begin{equation}
\frac{\bf k\cdot{\bm \nabla}}{m}\delta\rho^{qs}_{{\bf
    k}0}=-\theta\left(\frac{kv_s}{E_k}\right)\frac{\xi_k}{E_k\tau_k}|\Delta|^2({\bf
  k}\cdot{\bf v_s})\frac{m^+_{\bf k}-m^-_{\bf k}}{2},
\end{equation}
from which, one obtains
\begin{equation}
\delta\rho^{qs}_{{\bf
    k}}=|\Delta|^2({\bf
  k}\cdot{\bf v_s})\left(\begin{array}{cc}
\delta{m^+_{\bf k}}& 0\\
0 &\delta{m^-_{\bf k}} 
\end{array}\right),
\end{equation}
with $\delta{m}^{\pm}_{\bf
  k}=\mp\frac{\xi}{\tau_kv_F}\theta(\frac{kv_s}{E_k})\frac{\xi_k}{E_k}m^{\pm}_{\bf
  k}$. Here, we have used ${|\Delta|^2}/({{\nabla}|\Delta|^2})=\xi$.
  Consequently, Eq.~(\ref{mr}) is obtained.

\section{Derivation of Ginzburg-Landau equation}
\label{ac}

In this part, we derive the Ginzburg-Landau equation. By using Eq.~(\ref{E7})
to substitute $\rho_{{\bf k}+}$ into the gap equation [Eq.~(\ref{ge})], one has
\begin{eqnarray}
&&-\frac{\Delta}{V}={\sum_{\bf k}}'\bigg[\frac{\rho_{{\bf k}3}}{\xi_k}-\frac{i{\bm \partial}_{\bf k}\rho_{{\bf
    k}0}\cdot({\bm \nabla}-2ie{\bf A})}{2\xi_k}-\frac{\varepsilon_{{\bf p}-2e{\bf
A}}}{4\xi_k\Delta}\rho_{{\bf k}+}\nonumber\\
&&\mbox{}-\frac{{\bm \partial}_{\bf k}{\bm \partial}_{\bf k}\rho_{{\bf
    k}3}:({\bm \nabla}-2ie{\bf A})({\bm \nabla}-2ie{\bf
  A})}{8\xi_k}\bigg]\Delta.\label{ac1}
\end{eqnarray}
At $v_s<v_L$ with only the non-viscous superfluid, the superconducting state behaves
like the BCS one. Consequently, the density matrix in the quasiparticle space
reads 
\begin{equation}
\rho^{q}_{\bf k}=\left(\begin{array}{cc}
f(E_k)& 0\\
0 &1-f(E_k) 
\end{array}\right).\label{ac2}
\end{equation} 
With this BCS-state density matrix in the quasiparticle space, by treating
$\Delta$ as a small quantity near the critical
temperature, Eq.~(\ref{ac1}) becomes
\begin{eqnarray}
&&\frac{\Delta}{D_0V}=\int^{\omega_D}_{-\omega_D}{d}\xi_k\bigg[{\partial}^2_{k}\left(\frac{\xi_k\rho^q_{{\bf
    k}3}}{E_k}\right)\frac{({\bm \nabla}-2ie{\bf
  A})^2}{24\xi_k}-\frac{\rho^q_{{\bf k}3}}{E_k}\bigg]\Delta\nonumber\\
&&\approx\int^{\omega_D}_{-\omega_D}{d}\xi_k\Bigg\{\frac{1-2f(|\xi_k|)}{2|\xi_k|}+\frac{|\Delta|^2\partial_{|\xi_k|}}{2|\xi_k|}\bigg[\frac{1-2f(|\xi_k|)}{2|\xi_k|}\bigg]\nonumber\\
&&\mbox{}+\bigg(\frac{2\partial_{E_k}}{E_k}+\partial^2_{E_k}\bigg)\left(\frac{2f(E_k)-1}{2E_k}\right)\frac{k^2_F}{2m}\frac{({\bm \nabla}-2ie{\bf
  A})^2}{12m}\Bigg\}\Delta.\nonumber\\ \label{ac3}
\end{eqnarray}
With
$\frac{1}{D_0V}=\ln\Big(\frac{2\gamma}{\pi}\frac{\omega_D}{T_c}\Big)$ in the
BCS theory,\cite{BCS} from Eq.~(\ref{ac3}), one obtains
\begin{equation}
\label{GL}
\Bigg\{\frac{({\bm
  \nabla}-2ie{\bf A})^2}{4m}+\frac{1}{\lambda}\left[\alpha-\beta|\Delta|^2\right]\Bigg\}\Delta=0,
\end{equation}
with 
\begin{eqnarray}
\alpha&=&\int^{\omega_D}_{-\omega_D}{d}\xi_k\frac{1-2f(|\xi_k|)}{2|\xi_k|}-\frac{1}{D_0V}\nonumber\\
&=&\ln\Big(\frac{2\gamma}{\pi}\frac{\omega_D}{T}\Big)-\ln\Big(\frac{2\gamma}{\pi}\frac{\omega_D}{T_c}\Big)=\ln\left(\frac{T}{T_c}\right),\\
\beta&=&\int^{\omega_D}_{-\omega_D}{d}\xi_k\frac{1}{2|\xi_k|}\partial_{|\xi_k|}\bigg[\frac{2f(|\xi_k|)-1}{2|\xi_k|}\bigg]\nonumber\\
&=&T\sum_{n}\int^{\omega_D}_{-\omega_D}{d}\xi_k\frac{1}{2|\xi_k|}\partial_{|\xi_k|}\left[\frac{1}{(i\omega_n)^2-\xi_k^2}\right]\nonumber\\
&\approx&T\sum_{n}\int^{\infty}_{-\infty}{d}\xi_k\frac{1}{[(\omega_n)^2+\xi_k^2]^2}=\frac{7R(3)}{8(\pi{T})^2},\\
\lambda&=&\frac{\varepsilon_{k_F}}{3}\int^{\omega_D}_{-\omega_D}{d}\xi_k\bigg(\frac{2}{E_k}+\partial_{E_k}\bigg)\partial_{E_k}\left(\frac{2f(E_k)-1}{2E_k}\right)\nonumber\\
&\approx&\frac{\varepsilon_{k_F}}{3}\int^{\infty}_{-\infty}{d}\xi_k\frac{\partial^2_{|\xi_k|}f(|\xi_k|)}{|\xi_k|}\nonumber\\
&=&T\sum_n\frac{2\varepsilon_{k_F}}{3}\int^{\infty}_{-\infty}{d}\xi_k\frac{1}{(i\omega_n-|\xi_k|)^3|\xi_k|}\nonumber\\
&=&T\sum_{n>0}\frac{4\varepsilon_{k_F}}{3}\int^{\infty}_{-\infty}{d}\xi_k\frac{3\omega_n^2-\xi_k^2}{(\xi^2_k+\omega^2_n)^3}=\varepsilon_{k_F}\frac{7R(3)}{6(\pi{T})^2}.\nonumber\\
\end{eqnarray}
Here, $\omega_n=(2n+1)\pi{T}$ represents the Matsubara frequency.\cite{G1}
Consequently, the Ginzburg-Landau equation\cite{Ginzburg,Gor-G,G1} is exactly
derived in Eq.~(\ref{GL}).

\section{Derivation of Eq.~(\ref{or})}
\label{ad}

We give the derivation of Eq.~(\ref{or}) in this section.
Following the derivation of the density matrix in the magnetic response, in the
optical response, at the weak scattering limit, 
the solution of density matrix reads
\begin{equation}
\rho^q_{\bf k}=\rho^{q0}_{\bf k}+\delta\rho^q_{\bf k}+\delta\rho^{qs}_{\bf k},\label{ad1}
\end{equation}
with
\begin{equation}
\rho^{q0}_{\bf k}=\left(\begin{array}{cc}
f(E^+_{\bf k})& 0\\
0 &f(E^-_{\bf k}) 
\end{array}\right).
\end{equation} 
Here, $\rho^{q0}_{\bf k}$ is the quasiparticle distribution of the FFLO-like
state with $E^{\pm}_{\bf k}={\bf k}\cdot{\bf v}_s{\pm}E_k$; $\delta\rho^q_{\bf
  k}$ denotes the disturbance from the FFLO-like state in the optical response
in the absence of the scattering; $\delta\rho^{qs}_{\bf k}$ represents the scattering
contribution.  

In Eq.~(\ref{O2}), one has {\small{$t_3=(u^2_k-v_k^2)\tau_3-2u_kv_k\tau_1$}},
 {\small{$t_2=\tau_2$}},  {\small{$t_1=(u^2_k-v_k^2)\tau_1+2u_kv_k\tau_3$}} and  {\small{$U^{\dagger}_k{\bm \partial}_{\bf
  k}U_k=\frac{i}{2}\frac{\bf k}{m}\frac{|\Delta|}{E_k^2}\tau_2$}} as well as  {\small{$U^{\dagger}_k{\bm \partial}_{\bf
  k}{\bm \partial}_{\bf
  k}U_k=-\frac{\bf k}{m}\frac{\bf
  k}{m}\Big(\frac{|\Delta|^2}{4E_k^4}\tau_0-ib_k\tau_2\Big)$}} with {\small{$b_k=\frac{\Delta}{4\varepsilon_kE^2_k}-\frac{\xi_k\Delta}{E^4_k}$}}.
 As revealed in the previous
 work,\cite{GOBE1,GOBE2,GOBE3} in the optical response,      
 the effective chemical potential $\mu_{\rm
   eff}$, determined from the charge neutrality condition,\cite{GOBE1} is excited and
 then involved in the kinetic equation as a feedback. Considering the large
 Fermi energy in conventional superconductors, for a relatively weak optical field, we
 neglect the feedback of $\mu_{\rm eff}$ in Eq.~(\ref{O2}).  
Then, by substituting the density matrix [Eq.~(\ref{ab1})] into Eq.~(\ref{O2}), one can
construct $\delta\rho^q_{\bf k}$ as  
\begin{equation}
\delta\rho^q_{\bf k}=-({\bf k}{\cdot}{\bf v}_s)\rho^{q1}_{\bf
  k}+\frac{({\bf k}{\cdot}{\bf v}_s)^2}{2}\rho^{q2}_{\bf
  k}+mv^2_s\rho^{q3}_{\bf k},\label{ad2}
\end{equation}
from which, the linear-order terms of $({\bf k}\cdot{\bf v}_s)$ in Eq.~(\ref{O2}) becomes
\begin{eqnarray}
&&i({\bf k}\cdot{\bf v}_s)(\omega\rho^{q1}_{\bf k}+2E_k\rho^{q1}_{{\bf
    k}+}\tau_+-2E_k\rho^{q1}_{{\bf k}-}\tau_-)=e({\bf k}{\cdot}{\bf E})\frac{|\Delta|^2}{E^2_k}\nonumber\\
&&\mbox{}\times\left(\frac{1}{E_k}-\partial_{E_k}\right)\rho^{q0}_{{\bf k}3}\tau_0+{\bf
  k}\cdot\left(i\omega{\bf v}_s+\frac{e{\bf
      E}}{m}\right)\partial_{E_k}\rho^{q0}_{{\bf k}3}\tau_0. \nonumber\\ \label{ad3}
\end{eqnarray}
Here, we have neglected $\partial_{E_k}\rho^{q0}_{{\bf k}0}$ terms,  which is zero in
either pairing or unpairing regions. From Eq.~(\ref{ad3}), one has ${\bf
  v}_s=-\frac{e{\bf E}}{i\omega{m}}$ and $\rho^{q1}_{\bf k}=\frac{\rho_{m{\bf
      k}}\tau_0}{4\varepsilon_{k_F}}$. 

By using $e{\bf E}/m=-i\omega{\bf v_s}$, the nonlinear-order terms of $({\bf k}\cdot{\bf
  v}_s)^2$ in Eq.~(\ref{O2}) reads
\begin{eqnarray}
&&i({\bf k}\cdot{\bf v}_s)^2(\omega\rho^{q2}_{\bf k}+E_k\rho^{q2}_{{\bf
    k}+}\tau_+-E_k\rho^{q2}_{{\bf k}-}\tau_-)=i({\bf k}\cdot{\bf v}_s)^2\nonumber\\
&&\mbox{}\times\sum_{j=\pm}\Bigg[\omega\frac{\xi^2_k}{2E^2_k}\partial_{E_k}\rho^{q1}_{{\bf
    k}0}\tau_3-(\omega+2jE_k)\frac{\xi_k|\Delta|}{E^2_k}\partial_{E_k}\rho^{q1}_{{\bf
    k}0}\tau_{j}\nonumber\\
&&\mbox{}+j\tau_j\frac{|\Delta|}{E^2_k}\bigg(\xi_k^2\partial^2_{E_k}\rho^{q0}_{{\bf
  k}3}-3\xi_k\rho^{q1}_{{\bf
    k}0}+E^2_k\frac{\partial_{E_k}\rho^{q0}_{{\bf
  k}3}+\rho^{q1}_{{\bf
  k}0}}{2\varepsilon_{k}}\bigg)\Bigg], \nonumber\\ \label{ad6}
\end{eqnarray}
from which, one has
\begin{eqnarray}
&&\rho^{q2}_{\bf k}=\frac{\xi^2_k}{E^2_k}\partial_{E_k}\rho^{q1}_{{\bf
    k}0}\tau_3-\sum_{j=\pm}\Bigg[\frac{\omega+2jE_k}{\omega+jE_k}\frac{\xi_k|\Delta|}{E^2_k}\partial_{E_k}\rho^{q1}_{{\bf
    k}0}\tau_{j}\nonumber\\
&&\mbox{}+\frac{j|\Delta|\tau_j}{\omega+jE_k}\bigg(\frac{\xi_k^2\partial^2_{E_k}\rho^{q0}_{{\bf
  k}3}-3\xi_k\rho^{q1}_{{\bf
    k}0}}{E_k^2}+\frac{\partial_{E_k}\rho^{q0}_{{\bf
  k}3}+\rho^{q1}_{{\bf
  k}0}}{2\varepsilon_{k}}\bigg)\Bigg].\nonumber\\ \label{ad7}
\end{eqnarray}
For the nonlinear-order terms of $v^2_s$ in Eq.~(\ref{O2}), one obtains
\begin{eqnarray}
&&2i{v^2_s}(\omega\rho^{q3}_{\bf k}+E_k\rho^{q3}_{{\bf
    k}+}\tau_+-E_k\rho^{q3}_{{\bf k}-}\tau_-)=v^2_s\Bigg[\frac{|\Delta|}{E_k}\rho^{q0}_{{\bf
    k}3}\tau_2\nonumber\\
&&\mbox{}+i\omega\left(\frac{\xi_k}{E_k}\tau_3-\frac{|\Delta|}{E_k}\tau_1\right)\rho^{q1}_{{\bf
    k}0}+2\Delta\rho^{q1}_{{\bf k}0}\tau_2\Bigg], \label{ad4}
\end{eqnarray}
from which, $\rho^{q3}_{\bf k}$ reads
\begin{equation}
\rho^{q3}_{\bf k}=\frac{\xi_k}{2E_k}\rho^{q1}_{{\bf k}0}\tau_3-\frac{|\Delta|}{2E_k}\sum_{j=\pm}\frac{[j\rho^{q0}_{{\bf
    k}3}+\rho^{q1}_{{\bf
    k}0}(\omega+2jE_k)]\tau_{j}}{\omega+j{E_k}}.\label{ad5}
\end{equation}
As for the scattering contribution $\delta\rho^{qs}_{\bf k}$, one has
\begin{equation}
\partial_T\delta\rho^{qs}_{{\bf
    k}}=\left\{\partial_{t}\rho^q_{\bf k}\Big|_{\rm sc}\right\}.\label{ad8} 
\end{equation}
In the weak scattering limit, the scattering only causes the momentum (current)
relaxation. Therefore, by keeping the linear-order terms of $({\bf k}\cdot{\bf
  v}_s)$ in $\rho^q_{\bf k}$, from Eq.~(\ref{s5}), Eq.~(\ref{ad8}) becomes
\begin{equation}
\partial_T\delta\rho^{qs}_{{\bf
    k}0}=-\frac{({\bf k}\cdot{\bf v}_s)}{\tau_k}\theta\left(\frac{kv_s}{E_k}\right)(\partial_{E_k}\rho^{q0}_{{\bf k}3}+{\hat
  O_k}f_{\bf k}),
\end{equation}
where {\small{${\hat
      O_k}=4u_k^2v_k^2(1/{E_k}-\partial_{E_k})$}} and {\small{$f_{\bf k}=[3f(E^+_{\bf k})-3f(E^-_{\bf k})-f(E^+_{\bf
      k}+2E_k)+f(E^-_{\bf k}-2E_k)]/{8}$}}. Thus, $\delta\rho^{qs}_{{\bf
    k}}$ is obtained as 
\begin{equation}
\delta\rho^{qs}_{{\bf
    k}0}=-\frac{({\bf k}{\cdot}{\bf v}_s)}{i\omega\tau_k}\theta\left(\frac{kv_s}{E_k}\right)(\partial_{E_k}\rho^{q0}_{{\bf k}3}+{\hat
  O_k}f_{\bf k})\tau_0. 
\end{equation}
Consequently, Eq.~(\ref{or}) is obtained.

\begin{figure}[htb]
  {\includegraphics[width=8.0cm]{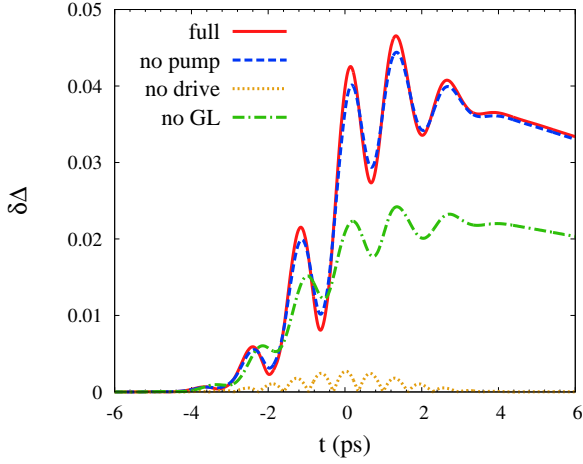}}
\caption{(Color online) Temporal evolutions of the Higgs mode from the numerical
  calculation of Eq.~(\ref{O1}). Blue dashed curve: without the
  pump effect by setting $p^2_s/(2m)=0$ in
  Eq.~(\ref{O1}); Brown dotted curve: without the
  drive effect by setting ${\bm \partial}_{\bf k}\rho_{\bf k}=0$ in
  Eq.~(\ref{O1}); Green chain curve: without the Ginzburg-Landau
  kinetic effect by removing the last two terms on the left-hand side of
  Eq.~(\ref{O1}). In the calculation, we used a THz linear-polarized
  optical pulse: ${\bf p_s}=(e/\omega)E_0{\bf
    e_x}\sin(\omega{t})\exp[-t^2/(2\sigma_t^2)]$ and $\partial_t{\bf
    p_s}=eE_0{\bf e_x}\cos(\omega{t})\exp[-t^2/(2\sigma_t^2)]$ with  $\sigma_t$
  being the width of the optical pulse. The
  used parameter\cite{GOBE1} in the calculation includes $m=0.067m_e$,
  $\Delta_0=0.8~$meV, $E_0=0.1~$kv/cm, $\sigma_t=2~$ps, $n_i=0.2N_0$, $\kappa=12.9$,
  $N_0=5\times10^{11}~$cm$^{-2}$, $V=0.1788~$eV$\cdot${nm}$^3$,
  $\omega=2\Delta_0$ and $T=0.02~$K. For comparison, we take the same
  scattering terms as those in the previous theory by Yu and Wu.\cite{GOBE1}
 } 
\label{figyw5}
\end{figure}

\section{Comparison between drive and pump effects}
\label{ae}

We compare the drive and pump effects in the excitation of the Higgs
mode by performing a numerical calculation of Eq.~(\ref{O1}) in the presence of
a THz linear-polarized optical pulse. As seen from the numerical results plotted
in Fig.~\ref{figyw5}, a plateau of the superconducting gap is observed after
THz pulse as a consequence of the thermal effect and the excitation of the
Higgs mode (red solid curve) is dominated  by the drive effect (blue dashed
curve) whereas the pump effect (brown dotted curve) is marginal, in consistency
with our analytical analysis in Sec.~\ref{Higgs}. 

In addition, we also calculate the case without the
Ginzburg-Landau kinetic effect [last two terms on the left-hand side of
Eq.~(\ref{O1})], which exactly reduces to the previous theory by Yu and Wu.\cite{GOBE1}  
As seen from Fig.~\ref{figyw5}, in comparison to the full results (red solid
curve), the absence of the Ginzburg-Landau kinetic effect, represented by green
chain curve, leads to a quantitative reduction in the excitation of the Higgs
mode. In order to compare the excitation of the Higgs mode between our
theory and Ref.~\onlinecite{GOBE1}, we separate the drive
effect as
\begin{equation}
  {\delta\Delta^{\rm drive}}={\delta\Delta^{\rm drive}_{\rm no~GL}}+{\delta\Delta^{\rm
    drive}_{\rm GL}},\label{full}
\end{equation}
with ${\delta\Delta^{\rm drive}_{\rm
    no~GL}}$ denoting the pure drive effect  [from forth term on the left-hand side of Eq.~(\ref{O1})] in the absence
of the Ginzburg-Landau kinetic effect and ${\delta\Delta^{\rm drive}_{\rm
    GL}}$ representing the contribution exactly from the Ginzburg-Landau kinetic effect. By using the same
technique in Appendix.~\ref{ad} to derive the nonlinear response, one has 
\begin{eqnarray}
\frac{\delta\Delta^{\rm
    drive}_{\rm GL}}{\Delta_0V}&=&\varepsilon_{p_s}\sum_{{\bf k}\in{\rm
    P}}\Bigg\{\bigg[2\varepsilon_{k_F}\frac{\xi^2_k\cos^2\theta_{\bf k}\partial^2_{E_k}f(E_k)}{E_k(\omega^2-E^2_k)}\nonumber\\
&&\mbox{}\times\left(1-\frac{\omega|\Delta_0|^2}{E^3_k}\right)+a_k\frac{|\Delta|^2}{E^3_k}\frac{E_k\cos^2{\theta_{\bf 
    k}}-\omega}{\omega^2-E^2_k}\bigg]\nonumber\\
&&\mbox{}-2\bigg[\frac{\varepsilon_{k_F}\cos\theta^2_{\bf
    k}\xi^2_k}{E_k(\omega^2-E^2_k)}\partial_{E_k}-\frac{\omega}{\omega^2-E_k^2}\bigg]\rho^{q1}_{{\bf
k}0},\nonumber \\ \label{ae1}\\
\frac{\delta\Delta^{\rm drive}_{\rm no~GL}}{\Delta_0V}&=&\varepsilon_{p_s}\sum_{{\bf k}\in{\rm
    P}}2\bigg[\frac{\varepsilon_{k_F}\cos\theta^2_{\bf
    k}\xi^2_k}{E_k(\omega^2-E^2_k)}\partial_{E_k}-\frac{\omega}{\omega^2-E_k^2}\bigg]\rho^{q1}_{{\bf
k}0}.\nonumber\\
\end{eqnarray}
As seen from above, in the complete contribution ${\delta\Delta^{\rm drive}}$ [Eq.~(\ref{full})],
the pure drive effect ${\delta\Delta^{\rm drive}_{\rm
    no~GL}}$ in the previous work\cite{GOBE1} is canceled by the second term in ${\delta\Delta^{\rm drive}_{\rm
    GL}}$ [Eq.~(\ref{ae1})], leaving only the contribution from the first term in ${\delta\Delta^{\rm drive}_{\rm
    GL}}$ [Eq.~(\ref{ae1})].

\end{appendix}


\begin{thebibliography}{0}

\bibitem{Ba0} J. R. Schrieffer, {\em Theory of Superconductivity} (W. A. Benjamin,
New York, 1964).
\bibitem{Ba1} M. Tinkham, {\em Introduction to Superconductivity} (McGraw-Hill,
New York, 1975).
\bibitem{Ba2} C. J. Pethick and H. Smith, J. Phys. C: Solid State Phys. {\bf
    13}, 6313 (1980).
\bibitem{Ba3} {\em Non-Equilibrium Superconductivity}, edited by D. N.
Langenderg and A. Larkin (North-Holland, Amsterdam, 1980).
\bibitem{Ba4} {\em Principles of Superconductive Devices and Circuits}, edited by T. Van Duzer and C. W. Turner
(North-Holland, Amsterdam, 1981). 
\bibitem{Ba5} N. Kopnin, {\em Theory of Nonequilibrium Superconductivity}
  (Oxford University Press, New York, 2001). 
\bibitem{Ba6} V. Chandrasekhar, in {\em The Physics of Superconductors}, edited
by K. H. Bennemann and J. B. Ketterson (Springer, Berlin, 2004), Vol. II.
\bibitem{Ba7} F. S. Bergeret, A. F. Volkov, and K. B. Efetov, Rev. Mod. Phys.
{\bf 77}, 1321 (2005).
\bibitem{Ba8} A. I. Buzdin, Rev. Mod. Phys. {\bf 77}, 935 (2005).

\bibitem{Meissner} W. Meissner and R. Ochsenfeld, Naturwissenschaften {\bf 21},
  787 (1933). 
\bibitem{London}  F. London and H. London, Proc. R. Soc. A {\bf 149}, 71 (1935).
\bibitem{Ginzburg} V. L. Ginzburg and L. D. Landau, Zh. Eksperim. i
  Teor. Fiz. {\bf 20}, 1064 (1950).

\bibitem{L1}  N. M. Rugheimer, A. Lehoczky, and C. V. Briscoe,
Phys. Rev. {\bf 154}, 414 (1967).
\bibitem{L2} S. L. Norman, Phys. Rev. {\bf 167}, 393 (1968).
\bibitem{L3} D. R. Karecki, G. L. Carr, S. Perkowitz, D. U. Gubser, and S. A. Wolf,
Phys. Rev. B {\bf 27}, 5460 (1983). 
\bibitem{L4}  D. E. Oates, A. C. Anderson, C. C. Chin, J. S. Derov,
  G. Dresselhaus, and M. S. Dresselhaus, Phys. Rev. B {\bf 43}, 7655 (1991). 
\bibitem{L5} J. F. Federici, B. I. Greene, P. N. Saeta, D. R. Dykaar, F. Sharifi, and R. C. Dynes,
Phys. Rev. B {\bf 46}, 11153 (1992). 
\bibitem{L6} S. D. Brorson, R. Buhleier, I. E. Trofimov, J. O. White, C. Ludwig,
  F. F. Balakirev, H. U. Habermeier, and J. Kuhl, J. Opt. Soc. Am. B {\bf 13},
  1979 (1996).  
\bibitem{L7} G. L. Carr, R. P. S. M. Lobo, J. LaVeigne, D. H. Reitze, and D. B. Tanner,
Phys. Rev. Lett. {\bf 85}, 3001 (2000). 
\bibitem{L8} A. V. Pronin, A. Pimenov, A. Loidl, and S. I. Krasnosvobodtsev,
  Phys. Rev. Lett. {\bf 87}, 097003 (2001).
\bibitem{L9} Z. Dai and P. A. Lee, Phys. Rev. B {\bf 95}, 014506 (2017).

\bibitem{NL1} R. A. Kaindl, M. A. Carnahan, J. Orenstein, D. S. Chemla, H.
M. Christen, H. Y. Zhai, M. Paranthaman, and D. H. Lowndes,
Phys. Rev. Lett. {\bf 88}, 027003 (2001). 
\bibitem{NL2} J. Demsar, R. D. Averitt, A. J. Taylor, V. V. Kabanov, W. N.
Kang, H. J. Kim, E. M. Choi, and S. I. Lee, Phys. Rev. Lett. {\bf 91}, 267002
(2003). 
\bibitem{NL3} R. A. Kaindl, M. A. Carnahan, D. S. Chemla, S. Oh, and J. N.
Eckstein, Phys. Rev. B {\bf 72}, 060510(R) (2005).
\bibitem{NL4} R. Kaneko, I. Kawayama, H. Murakami, and M. Tonouchi,
Appl. Phys. Express {\bf 3}, 042701 (2010).
\bibitem{NL5} M. Beck, M. Klammer, S. Lang, P. Leiderer, V. V. Kabanov,
G. N. Gol'tsman, and J. Demsar, Phys. Rev. Lett. {\bf 107}, 177007 (2011).
\bibitem{NL6} A. Glossner, C. Zhang, S. Kikuta, I. Kawayama, H. Murakami,
P. M{\" u}ller, and M. Tonouchi, arXiv:1205.1684. 
\bibitem{NL7} R. Matsunaga and R. Shimano, Phys. Rev. Lett. {\bf 109}, 187002
(2012).
\bibitem{NL8} R. Matsunaga, Y. I. Hamada, K. Makise, Y. Uzawa, H. Terai, Z.
Wang, and R. Shimano, Phys. Rev. Lett. {\bf 111}, 057002 (2013).
\bibitem{NL9} R. Matsunaga, N. Tsuji, H. Fujita, A. Sugioka, K. Makise, Y.
Uzawa, H. Terai, Z. Wang, H. Aoki, and R. Shimano, Science {\bf 345}, 1145 (2014).
\bibitem{NL10} R. Matsunaga, N. Tsuji, K. Makise, H. Terai, H. Aoki, and
  R. Shimano, Phys. Rev. B {\bf 96}, 020505 (2017). 

\bibitem{Tisza} L. Tisza, C. R. Acad. Sci. {\bf 207}, 1035 (1938); {\bf 207}, 1186 (1938);
F. London, Phys. Rev. {\bf 54}, 947 (1938).
\bibitem{Landau} L. D. Landau, Zh. Eksp. Teor. Fiz. {\bf 11}, 592 (1941);
  J. Phys. USSR {\bf 5}, 71 (1941).
\bibitem{two} J. Bardeen, Phys. Rev. Lett. {\bf 1}, 399 (1958).

\bibitem{Higgs1} R. A. Barankov, L. S. Levitov, and B. Z. Spivak,
  Phys. Rev. Lett. {\bf 93}, 160401 (2004).
\bibitem{Higgs2} R. A. Barankov and L. S. Levitov, Phys. Rev. Lett. {\bf 96},
  230403 (2006). 
\bibitem{Higgs20} T. Papenkort, V. M. Axt, and T. Kuhn, Phys. Rev. B {\bf 76}, 224522 (2007). 
\bibitem{Higgs21} T. Papenkort, T. Kuhn, and V. M. Axt, Phys. Rev. B {\bf 78},
  132505 (2008).  
\bibitem{Higgs3} A. Moor, P. A. Volkov, A. F. Volkov, and K. B. Efetov, Phys.
Rev. B {\bf 90}, 024511 (2014).
\bibitem{Higgs4} N. Tsuji and H. Aoki, Phys. Rev. B {\bf 92}, 064508 (2015). 
\bibitem{Higgs5} A. F. Kemper, M. A. Sentef, B. Moritz, J. K. Freericks, and
  T. P. Devereaux, Phys. Rev. B {\bf 92}, 224517 (2015). 
\bibitem{Higgs50} M. Dzero, M. Khodas, and A. Levchenko, Phys. Rev. B {\bf 91}, 214505 (2015).
\bibitem{Higgs6} H. Krull, N. Bittner, G. S. Uhrig, D. Manske, and A. P.
Schnyder, Nat. Commun. {\bf 7}, 11921 (2016).
\bibitem{Higgs7} M. Lu, H. W. Liu, P. Wang, and X. C. Xie, Phys. Rev. B {\bf 93},
064516 (2016).
\bibitem{Higgs8} N. Tsuji, Y. Murakami, and H. Aoki, Phys. Rev. B {\bf 94},
  224519 (2016).
\bibitem{Higgs9} T. Cea, C. Castellani, and L. Benfatto, Phys. Rev. B {\bf 93},
  180507(R) (2016).
\bibitem{Higgs10} Y. Murotani, N. Tsuji, and H. Aoki, Phys. Rev. B {\bf 95}, 104503
  (2017).

\bibitem{BCS} J. Bardeen, L. N. Cooper, and J. R. Schrieffer, Phys. Rev. {\bf
    106}, 162 (1957).
 
\bibitem{MB} D. C. Mattis and J. Bardeen, Phys. Rev. {\bf 111}, 412 (1958).
\bibitem{MBm} P. B. Miller, Phys. Rev. {\bf 113}, 1208 (1958); {\bf 118}, 928
  (1960). 
\bibitem{Gor-A} A. A. Abrikosov, L. P. Gor'kov, Zh. Exp. Teor. Fiz. {\bf 35},
  1558 (1958) [Sov. Phys. JETP {\bf 8}, 1090 (1959)];  Zh. Exp. Teor. Fiz. {\bf 36},
  319 (1959) [Sov. Phys. JETP {\bf 9}, 220 (1959)]. 
\bibitem{Gor-o} A. A. Abrikosov, L. P. Gor'kov, I. M. Khalatnikov,
  Zh. Eksp. Teor. Fiz. {\bf 35}, 265 (1958) [Sov. Phys. JETP {\bf 8}, 182
  (1959)]. 
\bibitem{Gor-G}  L. P. Gor'kov, Zh. Eksp. Teor. Fiz. {\bf 36}, 1918 (1959)
  [Sov. Phys. JETP {\bf 9}, 1364 (1959)]. 
\bibitem{G1} A. A. Abrikosov, L. P. Gor'kov, and I. E. Dzyaloshinski, {\em Methods of Quantum Field Theory
in Statistical Physics} (Prentice Hall, Englewood Cliffs, 1963). 
\bibitem{Gor-oMB} G. Rickayzen, {\em Theory of Superconductivity} (John Wiley \&
  Sons, Inc., New York, 1965). 
\bibitem{G2} L. P. Gor'kov and G. M. Eliashberg, Zh. Eksp. Teor. Fiz. {\bf 51}, 612 (1968) [Sov. Phys. JETP
{\bf 27}, 328 (1968)].
\bibitem{Eilen} G. Eilenberger, Z. Phys. {\bf 214}, 195 (1968).
\bibitem{Usadel} K. D. Usadel, Phys. Rev. Lett. {\bf 25}, 507 (1970).
\bibitem{MBo} S. B. Nam, Phys. Rev. {\bf 156}, 470 (1967); I. S. B. Nam, Phys.
Rev. B {\bf 2}, 3812 (1970).
\bibitem{GOBE1} T. Yu and M. W. Wu,  Phys. Rev. B {\bf 96}, 155311 (2017).
\bibitem{GOBE2} T. Yu and M. W. Wu,  Phys. Rev. B {\bf 96}, 155312 (2017).
\bibitem{GOBE3} F. Yang, T. Yu, and M. W. Wu,  Phys. Rev. B {\bf 97}, 205301
  (2018).  

\bibitem{gi0} Y. Nambu, Phys. Rev. {\bf 117}, 648 (1960).
\bibitem{gi1} V. Ambegaokar and L. P. Kadanoff, Il Nuovo Cimento {\bf 22}, 914 (1961).
\bibitem{gi2} Y. Nambu, Rev. Mod. Phys. {\bf 81}, 1015 (2009).


\bibitem{E1} G. E. Peabody and R. Meservey, Phys. Rev. B {\bf 6}, 2579 (1972).
\bibitem{E2} C. Varmazis and M. Strongin, Phys. Rev. B {\bf 10}, 1885 (1974).
\bibitem{E3} F. Behroozi, M. P. Garfunkel, F. H. Rogan, and G. A. Wilkinson,
Phys. Rev. B {\bf 10}, 2756 (1974).
\bibitem{E4} A. I. Gubin, K. S. Il'in, S. A. Vitusevich, M. Siegel, and N. Klein,
Phys. Rev. B {\bf 72}, 064503 (2005).
\bibitem{E5} T. R. Lemberger, I. Hetel, J. W. Knepper, and F. Y. Yang,
Phys. Rev. B {\bf 76}, 094515 (2007).

\bibitem{As} P. W. Anderson, Phys. Rev. {\bf 112}, 1900 (1958).

\bibitem{GQ1} J. Rammer and H. Smith, Rev. Mod. Phys. {\bf 58}, 323 (1986).
  
\bibitem{Eilen2Ginz} T. Kita, {\em Statistical Mechanics of Superconductivity}
  (Springer, Berlin, 2015).
\bibitem{EG1} T. Kita, Phys. Rev. B {\bf 64}, 054503 (2001).
\bibitem{EG2} F. Konschelle, Eur. Phys. J. B {\bf 87}, 119 (2014).
\bibitem{Wilson} M. E. Peskin and D. V. Schroeder, {\em An Introduction to Quantum
Field Theory} (Addison-Wesley, New York, 1995).
\bibitem{EJ1} F. S. Bergeret, A. F. Volkov, and K. B. Efetov, Phys. Rev. B {\bf
    64}, 134506 (2001).
\bibitem{EJ2} A. A. Golubov, M. Yu. Kupriyanov, and E. Il’ichev,
  Rev. Mod. Phys. {\bf 76}, 411 (2004).
\bibitem{EJ3} A. Buzdin, Phys. Rev. Lett. {\bf 101}, 107005 (2008).
  
\bibitem{EU1} A. B. Vorontsov, J. A. Sauls, and M. J. Graf, Phys. Rev. B {\bf
    72}, 184501 (2005). 
\bibitem{EU2} N. Hayashi, K. Wakabayashi, P. A. Frigeri, and M. Sigrist,
  Phys. Rev. B {\bf 73}, 024504 (2006). 
\bibitem{EU3} A. B. Vorontsov and I. Vekhter, Phys. Rev. B {\bf 75}, 224501 (2007).
\bibitem{EU4} K. An, T. Sakakibara, R. Settai, Y. Onuki, M. Hiragi, M. Ichioka,
  and K. Machida, Phys. Rev. Lett. {\bf 104}, 037002 (2010).
  
\bibitem{EV1} M. Ichioka, A. Hasegawa, and K. Machida, Phys. Rev. B {\bf 59},
  8902 (1999).
\bibitem{EV2} N. Nakai, P. Miranovi{\' c}, M. Ichioka, and K. Machida,
Phys. Rev. B {\bf 70}, 100503(R) (2004).
\bibitem{EV3} K. Watanabe, T. Kita, and M. Arai, Phys. Rev. B {\bf 71}, 144515
  (2005).
\bibitem{EV4} M. Ichioka and K. Machida, Phys. Rev. B {\bf 76}, 064502 (2007).
  
\bibitem{ED1} M. Houzet and V. P. Mineev, Phys. Rev. B {\bf 74}, 144522 (2006).
\bibitem{ED2} M. G. Vavilov and A. V. Chubukov, Phys. Rev. B {\bf 84}, 214521 (2011).
\bibitem{ED3} F. P. J. Lin and A. Gurevich, Phys. Rev. B {\bf 85}, 054513 (2012).
\bibitem{ED4} M. Hoyer, S. V. Syzranov, and J. Schmalian, Phys. Rev. B {\bf 89},
  214504 (2014).

\bibitem{U1} A. Schmid and G. Sch{\" o}n, J. Low Temp. Phys. {\bf 20}, 207 (1975).
\bibitem{U2}  A. L. Shelankov, Zh. Eksp. Teor. Fiz. {\bf 78}, 2359 (1980)
  [Sov. Phys. JETP {\bf 51}, 1186 (1980)]; J. Low Temp. Phys. {\bf 60}, 29
  (1985). 
\bibitem{U3}  U. Eckern, J. Low Temp. Phys. {\bf 50}, 489 (1983).
\bibitem{U4}  M. Y. Kuprianov and V. F. Lukichev, Zh. Eksp. Teor. Fiz. {\bf 94}, 139 (1988) [Sov. Phys.
JETP {\bf 67}, 1163 (1988)].
\bibitem{U5}  Y. Takane, J. Phys. Soc. Jpn. {\bf 75}, 074711 (2006).
\bibitem{U6}  F. S. Bergeret and I. V. Tokatly, Phys. Rev. Lett. {\bf 110},
  117003 (2013); Phys. Rev. B {\bf 89}, 134517 (2014).

\bibitem{GQ2} H. Haug and A. P. Jauho, {\em Quantum Kinetics in Transport and
Optics of Semiconductors} (Springer, Berlin, 1996).
\bibitem{GQ3} M. W. Wu, J. H. Jiang, and M. Q. Weng, Phys. Rep. {\bf 493}, 61
(2010).

\bibitem{FF4} I. Khavkine, H. Y. Kee, and K. Maki, Phys. Rev. B {\bf 70}, 184521 (2004).
\bibitem{FF5} G. Tkachov and V. I. Fal'ko, Phys. Rev. B {\bf 69}, 092503 (2004).
\bibitem{FF6} F. Rohlfing, G. Tkachov, F. Otto, K. Richter, D. Weiss, G. Borghs,
  and C. Strunk, Phys. Rev. B {\bf 80}, 220507(R) (2009). 
\bibitem{FF7} T. Yu and M. W. Wu, Phys. Rev. B {\bf 94}, 205305 (2016).
\bibitem{FF8} F. Yang and M. W. Wu, Phys. Rev. B {\bf 95}, 075304 (2017).
\bibitem{FF9} F. Yang and M. W. Wu, J. Low Temp. Phys. {\bf 192}, 241 (2018).

\bibitem{FF1} P. Fulde and R. A. Ferrell, Phys. Rev. {\bf 135}, 550 (1964).
\bibitem{FF2} A. I. Larkin and Y. N. Ovchinnikov, Zh. Eksp. Teor. Fiz. {\bf 47}, 1136
  (1964) [Sov. Phys. JETP {\bf 20}, 762 (1965)]. 
\bibitem{GKB} P. Lipavsk{\' y}, V. \v{S}pi\v{c}ka, and B. Velick{\' y},
  Phys. Rev. B {\bf 34}, 6933 (1986).
  
\bibitem{Pb1} G. I. Lykken,  A. L. Geiger,  K. S. Dy, and E. N. Mitchell,
  Phys. Rev. B  {\bf 4}, 1523 (1971).
\bibitem{Pb2} G. W. Webba, F. Marsigliob, and J. E. Hirsch, Physica C {\bf 514}, 17 (2015).


\bibitem{HTSC1} T. Timusk and B. Statt, Rep. Prog. Phys. {\bf 62}, 61
  (1999). 
\bibitem{HTSC2} S. Kleefisch, B. Welter, A. Marx, L. Alff, R. Gross, and
  M. Naito, Phys. Rev. B {\bf 63}, 100507(R) (2001). 
\bibitem{HTSC3} M. V. Sadovskii, Phys. Usp. {\bf 44}, 515 (2001).
\bibitem{HTSC4} P. A. Lee, N. Nagaosa, and X. G. Wen, Rev. Mod. Phys. {\bf
    78}, 17 (2006).
\bibitem{HTSC5} $\slashed{\rm O}$. Fischer, M. Kugler, I. M. Aprile, C. Berthod, and C. Renner,
Rev. Mod. Phys. {\bf 79}, 353 (2007). 
\bibitem{HTSC6} S. Badoux, W. Tabis, F. Lalibert{\'e}, G. Grissonnanche,
  B. Vignolle, D. Vignolles, J. B{\'e}ard, D. A. Bonn, W. N. Hardy, R. Liang,
  N. D. Leyraud, L. Taillefer, and C. Proust, Nature {\bf 531}, 210 (2016).  

\bibitem{SDSC1} M. Mondal, A. Kamlapure, M. Chand, G. Saraswat, S. Kumar,
  J. Jesudasan, L. Benfatto, V. Tripathi, and P. Raychaudhuri, 
Phys. Rev. Lett. {\bf 106}, 047001 (2011).
\bibitem{SDSC2} M. Chand, G. Saraswat, A. Kamlapure, M. Mondal, S. Kumar,
  J. Jesudasan, V. Bagwe, L. Benfatto, V. Tripathi, and P. Raychaudhuri,
  Phys. Rev. B {\bf 85}, 014508 (2012).
\bibitem{SDSC3} M. Mondal, A. Kamlapure, S. C. Ganguli, J. Jesudasan, V. Bagwe,
  L. Benfatto, and P. Raychaudhuri, Sci. Rep. {\bf 3}, 1357 (2013).
\bibitem{SDSC4} T. Dubouchet, B. Sac{\'e}p{\'e}, J. Seidemann, D. Shahar,
  M. Sanquer, and C. Chapelier, arXiv:1806.00323. 

\bibitem{STM} D. Eom, S. Qin, M. Y. Chou, and C. K. Shih, Phys. Rev. Lett. {\bf
    96}, 027005 (2006).
\bibitem{ARPES} A. Damascelli, Z. Hussain, and Z. X. Shen, Rev. Mod. Phys. {\bf
    75}, 473 (2003).
\bibitem{Drude}  P. Drude, Ann. Phys. (Leipzig) {\bf 1}, 566 (1900); {\bf 3}, 369
(1900).
  
\bibitem{cn0} Y. M. Galperin, V. L. Gurevich, V. I. Kozub, and A. L. Shelankov,
  Phys. Rev. B {\bf 65}, 064531 (2002). 
\bibitem{cn1} S. Takahashi and S. Maekawa, Phys. Rev. Lett. {\bf 88}, 116601
  (2002). 
\bibitem{cn2} S. Takahashi and S. Maekawa, J. Phys. Soc. Jpn. {\bf 77}, 031009 (2008).
\bibitem{cn3} S. Takahashi and S. Maekawa, Jpn. J. Appl. Phys. {\bf 51}, 010110
(2012).
\bibitem{cn4} H. L. Zhao and S. Hershfield, Phys. Rev. B {\bf 52}, 3632 (1995). 
\bibitem{cn5} S. Li, A. V. Andreev, and B. Z. Spivak, Phys. Rev. B {\bf 92},
  100506(R) (2015). 


\bibitem{cm1} R. Mersevery and B. B. Schwartz, {\em Superconductivity}, edited by
  R. D. Parks, (Marcel Dekker, New York, 1969). 
\bibitem{cm2} D. R. Tilley and J. Tilley, {\em Superﬂuidity and
    Superconductivity}, 2nd ed. (Adam Hilger, Bristol, 1986).
\bibitem{cm3} D. A. Bozhko, A. A. Serga, P. Clausen, V. I. Vasyuchka,
  F. Heussner, G. A. Melkov, A. Pomyalov, V. S. L'vov, and B. Hillebrands,
  Nat. Phys. {\bf 12}, 1057 (2016).
\bibitem{cm4} A. Moor, A. F. Volkov, and K. B. Efetov,  Phys. Rev. Lett. {\bf
    118}, 047001 (2017).

\bibitem{note} Actually, 
  the condensate should not be directly responsible for the supercurrent,
  since the Meissner supercurrent is proportional to the square of the superconducting
  order parameter whereas the charge density of the condensate ($2ev^2_k$) is not.
  
\bibitem{KB3} E. J. K{\" o}nig and A. Levchenko, Phys. Rev. Lett. {\bf 118}, 027001 (2017).


  
\end{thebibliography}
\end{document}